\documentclass[sigconf]{acmart}
\usepackage{amsmath}
\usepackage{color}
\usepackage{array}
\usepackage{breqn}
\usepackage{{inputenc}}
\usepackage{balance}
\usepackage{multirow,tabularx}
\usepackage{booktabs,longtable}
\usepackage{hhline}
\usepackage[flushleft]{threeparttable}
\usepackage{algorithm}
\usepackage{algorithmic}
\usepackage{epsfig}
\usepackage{graphicx}
\usepackage{epstopdf}
\usepackage{thmtools}
\usepackage{enumitem}
\usepackage{verbatim}
\usepackage{amsfonts}
\usepackage{hyperref}
\hypersetup{colorlinks=false,linkcolor=blue,urlcolor=blue,citecolor=red}
\usepackage{xcolor}
\usepackage{subcaption}
\usepackage{wrapfig}
\usepackage{makecell}

\newcommand{\Lapl}{\mathbf{\mathop{\mathcal{L}}}}

\newcommand{\Mat}[1]{\textbf{#1}}

\newcommand{\Space}[1]{\mathbb{#1}}
\newcommand{\Set}[1]{\mathcal{#1}}

\newcommand{\ie}{\emph{i.e., }}
\newcommand{\eg}{\emph{e.g., }}

\newcommand{\wrt}{\emph{w.r.t. }}

\newcommand{\za}[1]{{\color{black}{#1}}}
\newcommand{\jn}[1]{{\color{black}{#1}}}
\newcommand{\cmmnt}[1]{}

\DeclareMathOperator*{\argmax}{arg\,max}

\AtBeginDocument{%
  \providecommand\BibTeX{{%
    \normalfont B\kern-0.5em{\scshape i\kern-0.25em b}\kern-0.8em\TeX}}}

\copyrightyear{2023}
\acmYear{2023}
\setcopyright{rightsretained}
\acmConference[WWW '23]{Proceedings of the ACM Web Conference 2023}{April
30-May 4, 2023}{Austin, TX, USA}
\acmBooktitle{Proceedings of the ACM Web Conference 2023 (WWW '23), April
30-May 4, 2023, Austin, TX, USA}
\acmDOI{10.1145/3543507.3583461}
\acmISBN{978-1-4503-9416-1/23/04}

\begin{document}

\title{Invariant Collaborative Filtering to Popularity Distribution Shift}


\author{\textbf{An Zhang}}
\affiliation{%
  \institution{National University of Singapore}
  \institution{Sea-NExT Joint Lab}
  \country{}}
\email{anzhang@u.nus.edu}

\author{\textbf{Jingnan Zheng}}
\affiliation{%
  \institution{National University of Singapore}
  \country{}}
\email{e0718957@u.nus.edu}

\author{\textbf{Xiang Wang$^\S$}}\thanks{$^\S$Xiang Wang is the corresponding author, also with Institute of Artificial Intelligence, Hefei Comprehensive National Science Center.}
\affiliation{%
  \institution{University of Science and Technology of China}
  \country{}
  }
\email{xiangwang1223@gmail.com}

\author{\textbf{Yancheng Yuan}}
\affiliation{%
  \institution{The Hong Kong Polytechnic University}
  \country{}}
\email{yanchengyuanmath@gmail.com}

\author{\textbf{Tat-Seng Chua}}
\affiliation{%
  \institution{National University of Singapore}
  \institution{Sea-NExT Joint Lab}
  \country{}}
\email{dcscts@nus.edu.sg}


\begin{abstract}
Collaborative Filtering (CF) models, despite their great success, suffer from severe performance drops due to popularity distribution shifts, where these changes are ubiquitous and inevitable in real-world scenarios.
Unfortunately, most leading popularity debiasing strategies, rather than tackling the vulnerability of CF models to varying popularity distributions, require prior knowledge of the test distribution to identify the degree of bias and further learn the popularity-entangled representations to mitigate the bias.
Consequently, these models result in significant performance benefits in the target test set, while dramatically deviating the recommendation from users' true interests without knowing the popularity distribution in advance.
In this work, we propose a novel learning framework, \underline{Inv}ariant \underline{C}ollaborative \underline{F}iltering (\textbf{InvCF}), to discover disentangled representations that faithfully reveal the latent preference and popularity semantics without making any assumption about the popularity distribution.
At its core is the distillation of unbiased preference representations (\ie user preference on item property), which are invariant to the change of popularity semantics, while filtering out the popularity feature that is unstable or outdated. 
Extensive experiments on five benchmark datasets and four evaluation settings (\ie synthetic long-tail, unbiased, temporal split, and out-of-distribution evaluations) demonstrate that InvCF outperforms the state-of-the-art baselines in terms of popularity generalization ability on real recommendations.
Visualization studies shed light on the advantages of InvCF for disentangled representation learning. 
Our codes are available at \url{https://github.com/anzhang314/InvCF}.
\end{abstract}




\maketitle

\section{Introduction}

\begin{figure}[t]
    \centering
    \includegraphics[width=0.98\linewidth]{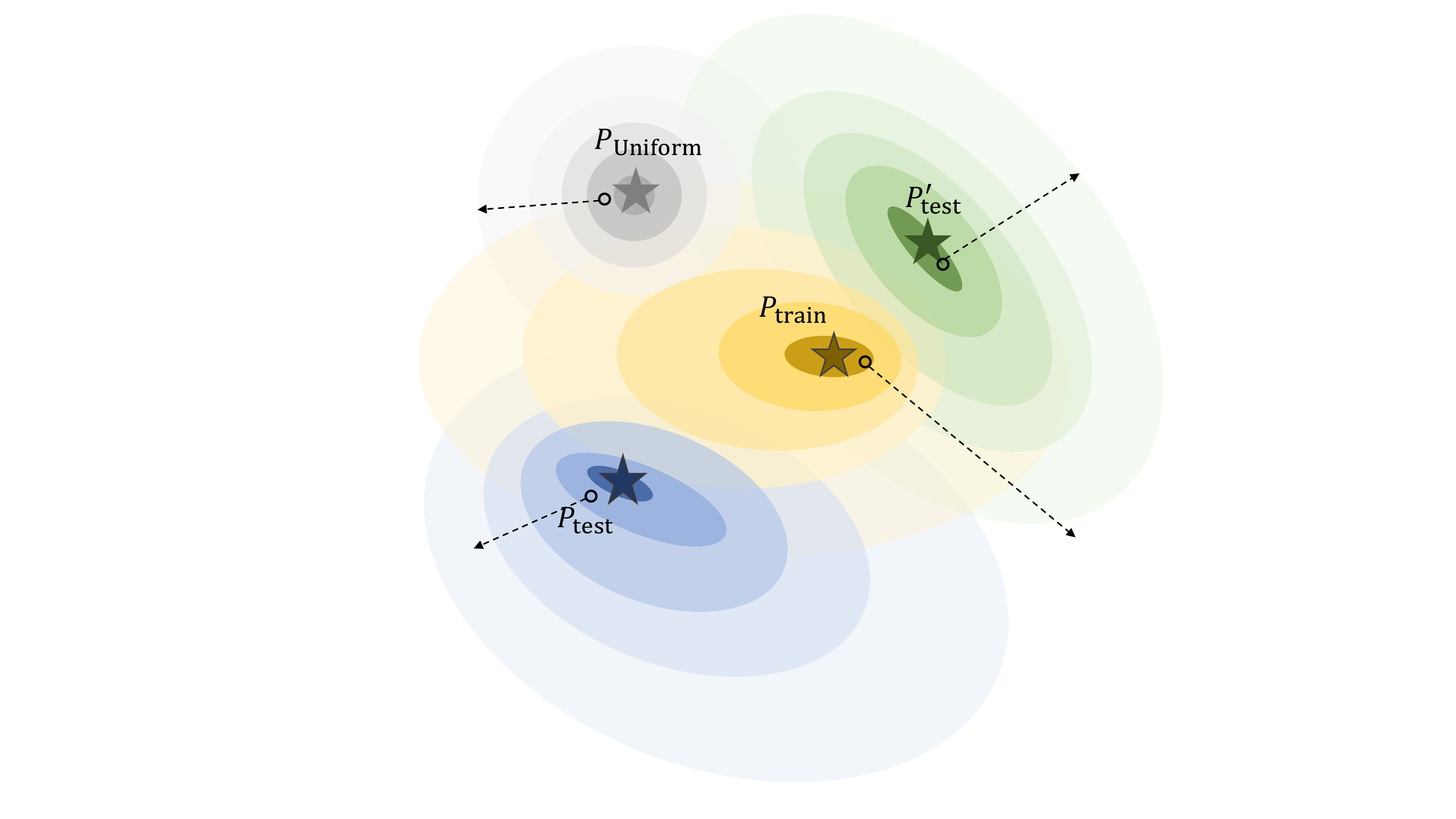}
    \caption{Illustration of real-world popularity distribution shifts.
    The alteration of the season naturally influences the item popularity distribution, where both $P_{test}$ and $P_{test}^'$ are possible to occur but are unpredictable.
    More datasets statistics can be found in Figure \ref{fig:yahoo} and Appendix \ref{sec:app_datasets}.}
    \label{fig:intro}
    \vspace{-15pt}
\end{figure}

Collaborative filtering (CF) is a keystone of personalized recommendation, hypothesizing that behaviorally similar users tend to have similar \jn{preferences} on items.
Inspecting leading CF models \cite{NGCF, KGAT,UltraGCN}, we can systematize a dominant paradigm --- view historical interactions between users and items as the training data, encode the collaborative signals as their representations, and then use these representations to predict future interactions in the test data.
It simply assumes that the training and test data are drawn from the same distribution.
However, this assumption hardly holds in the real-world scenarios, due to an inherent factor --- \textbf{popularity distribution shift} between the training and test data (See real-world datasets statistics in Figures \ref{fig:yahoo}, \ref{fig:coat}-\ref{fig:meituan}).
That is, the statistical popularity of items varies across historical and future interactions, which is usually caused by the demographic, regional, and chronological diversity of human behaviors \cite{CausPref,fairness}.
For example, the fashion trend and alteration of \jn{the} season will influence the change in item popularity (See Figure \ref{fig:intro}); meanwhile, the sales of take-out food will fluctuate between weekdays and weekends \cite{COR}.
Due to the prevalence of popularity distribution shifts, items and users representations are infused with unstable popularity correlations, which, under real-world recommendation scenarios, render CF models unrobust and severely degrade their performance.

To overcome the lack of robustness guarantees in real applications, popularity generalization over CF models is attracting a surge in interest \cite{bc_loss, Calibration, FPC, MostPop}.
However, rather than addressing the \jn{vulnerability} of CF model to popularity distribution shifts \cite{S-DRO,RankALS}, current prevalent studies mainly focus on popularity debiasing.
That is, measuring the degree of popularity bias in the training data first, and then mitigating the bias using a variety of debiasing strategies, including regularization-based \cite{ESAM, ALS+Reg, sam-reg, PC_Reg}, reweighting-based \cite{ Propensity_SVM-Rank, RecSys18_IPS, UBPR, BRD, DR, AutoDebias}, and causal embedding \cite{KDCRec, DICE, PDA, DecRS, CauSeR, 16_Causal_inference}.
Such the degree of bias is determined by the popularity distribution deviation of the training set from the target data.
Inevitably, to precisely quantify the deviation, there is an implicit but impractical constraint for these debiasing methods, \ie the target popularity distribution must be known in advance.
\begin{itemize}[leftmargin=*]
    \item The reweighting-based approaches \cite{ips, IPS-C, IPS-CN}, also referred \jn{to} as Inverse Propensity Score (IPS), inversely weight each item's prediction error with the propensity of its popularity.
    The propensities, once inverted, lead to an inherent assumption that the uniform popularity distribution with regard to items is unbiased and on target.
    \item Most cutting-edge debiasing techniques demand prior knowledge of the test popularity distributions, such as leveraging the validation set equipped with similar popularity distribution of test as a guide for hyperparameter adjustment \cite{DICE,MACR,BRD}, or leaking a small amount of unbiased data to strengthen unbiased recommendation learning \cite{KDCRec,CausE,AutoDebias}. 
\end{itemize}

These debiasing methods suffer from injecting popularity information into representations. 
As a result, without prior knowledge of the target distribution, highly popularity-entangled representations encounter a severe performance drop in practice, which further limits the applicability of these debiasing techniques.


We postulate that an ideal method to handle unknown popularity distribution shifts should learn disentangled representations that faithfully reflect the hidden invariant features and popularity semantics, rather than learning \jn{superficial} entangled representations.
Our core idea is, inspired by invariant learning \cite{IRM,REX} and disentangled representations \cite{disentangle}, to identify the invariant features (\ie a user's true preference, an item's real property) that causally determine the historical and future interactions, regardless of changes in variant popularity. 
Such popularity-invariant representations are able to yield a consistently plausible performance and enhanced generalization ability in real recommender systems (See Table \ref{tab:synthetic}).

Towards this end, we propose a new learning framework, \underline{Inv}ariant \underline{C}ollaborative \underline{F}iltering (\textbf{InvCF}), that integrates invariance and disentanglement principles.
By disentanglement principle, we mean that the representations are decomposed into two independent components - popularity and preference representations, while changing the ``popularity'' semantic does not affect the preference representations.
By invariance principle, we mean that the relations between preference representations and interactions are invariant throughout a variety of popularity distributions, and that preference representations are sufficient on their own to preserve the critical facts of interactions.

Guided by these two principles, our InvCF strategy incorporates four modules: a preference encoder, a popularity encoder, a representation augmentation module, and a representation disentangling module.
First, the popularity encoder and preference encoder, respectively, learn inference mappings from popularity statistics and historical interactions to the latent representation space.
Then the augmentation and disentangling modules are implemented based on the proposed principles to drive the representation learning.
Specifically, towards the disentanglement principle, disentangling module utilizes distance correlation as a regularizer to encourage independence of popularity and preference representations (See additional results employing various discrepancies in Table \ref{tab:discrepancy}). 
Towards the invariant principle, \jn{the} augmentation module concatenates the target preference \jn{representation} with other's popularity representations and enforces the \jn{prediction} to be invariant.
Jointly training under these two principles enables the CF models to disentangle the invariant/causal semantic features and variant/spurious popularity information, and further boosts the model's capability to popularity generalization. 
Our main contributions are summarized as follows:
\begin{itemize}[leftmargin=*]
    \item From a more realistic standpoint, we broaden our understanding of the current popularity debiasing problem and reformulate it as the problem of popularity distribution shift in CF.
    \item We propose a novel Invariant Collaborative Filtering (InvCF) method that hinges on the representation level disentanglement and augmentation to guarantee invariant feature learning.
    \item We conduct in-depth experiments with extensive test evaluations to justify the superiority of InvCF in diverse popularity distributions.
\end{itemize}



\section{Preliminary} \label{sec:prelimimary}
We begin with the definition of popularity distribution shift in CF, and highlight its differences from popularity bias.
Then, we formulate the problem of popularity generalization, and reveal the limitations of current debiasing approaches when facing this generalization problem.
Throughout the paper, we represent the random variables and their deterministic values with the upper-cased (\eg $X$) and lower-cased (\eg $x$) letters, respectively.

\subsection{Popularity Distribution Shift}
\noindent\textbf{Background.}
Here we focus on item recommendation from implicit feedback \cite{BPR}, where an interaction between a user and an item (\eg view, purchase, click) implicitly reflects the user's preference.
The task is building a CF model to learn the user preference from historical interactions and predict future interactions.
Let $\Set{D}_{\text{train}}=\{(x,y)|x=(u,i),y=1\}$ be the training set that involves historical interactions between users and items, and $\Set{D}_{\text{test}}=\{(x,y)|x=(u,i),y=1\}$ be the test set that contains future interactions, where $y=1$ indicates that user $u$ interacts with item $i$, otherwise $y=0$.

Formally, the dominant paradigm of learning CF models \cite{ImplicitFeedback,BPR} optimizes the model parameters $\hat{\theta}$ via maximum log-likelihood estimation on the training data $\Set{D}_{\text{train}}$:
\begin{gather}\label{equ:train-likelihood}
    \hat{\theta}=\argmax_{\theta}\log{P_{\text{train}}(Y|X)}=\argmax_{\theta}\sum_{(x,y)\in\Set{D}_{\text{train}}}\log{P}(y|x),
\end{gather}
where $X$ and $Y$ are the variables of user-item pair and interaction, respectively; $P(y|x)$ is the probability of $Y=y$ conditioned on $X=x$, indicating how likely user $u$ interacts with item $i$.
To approach the estimation, extensive model architectures \cite{BPR,LightGCN,UltraGCN,Multi-VAE} have been designed to develop the CF idea --- behaviorally similar users tend to have similar preferences on items.
Regardless of diverse designs, at the core is distilling the CF signals as the representation of $X$ and \za{regressing} it to $Y$.
For example, MF \cite{MF} and LightGCN \cite{LightGCN} learn a pair of user representation and item representation to depict $X$, and use the inner product of them to fit the interaction $Y$.

\vspace{5pt}
\noindent\textbf{Definition.}
Following prior studies \cite{DICE,CD2AN}, we reveal two parts inherent in $X$'s representation:
(1) information on users' pure \jn{preferences}, $Z_{\text{pref}}$, which reflects user interest in item properties;
(2) information on popularity, $Z_{\text{pop}}$, which describes user conformity influenced by item popularity.
On closer inspection on these parts \cite{InvPref,CausPref}, $Z_{\text{pref}}$ is more stable to serve as the causation of interaction $Y$; in stark contrast, $Z_{\text{pop}}$ more easily changes due to demographic, regional, and chronological diversity of human behaviors, thus holding the unstable correlation with interaction $Y$.
Conventional CF models mostly assume that the training and test data are from the same distribution, thereby having the same popularity information $Z_{\text{pop}}$. However, this assumption is unrealistic in the real-world scenarios.

Formally, across the training and test data (\ie $\Set{D}_{\text{train}}$ and $\Set{D}_{\text{test}}$), we define the underlying \jn{changes} in popularity information $Z_{\text{pop}}$ as popularity distribution shift:
\begin{gather}
    P_{\text{train}}(Z_{\text{pop}},Y)\neq P_{\text{test}}(Z_{\text{pop}},Y).
\end{gather}

\vspace{5pt}
\noindent\textbf{Differences from Popularity Bias.}
There has been increasing interest in the popularity distribution difference between the training and test data, which is quantitatively measured as popularity bias \cite{MACR,DICE,AutoDebias,ips}.
However, these studies on popularity bias inherently assume the popularity information of test distribution is known or assumed in advance during training.
See Section \ref{sec:preliminary-generalization} for more details.
In sheer contrast, \jn{the} popularity distribution shift focuses on the more general and practical scenario, which has no access to any popularity information about the test distribution.
That is, we hardly quantify the distribution discrepancy \wrt popularity between the training and test data.
Hence, it poses a major obstacle in enhancing the generalization ability of CF models to unknown popularity distribution shifts.

\subsection{Popularity Generalization}
\label{sec:preliminary-generalization}

\noindent\textbf{Problem Formulation.}
Following previous work \cite{OODSurvey}, we reformulate the recommendation problem with the focus on popularity generalization.
Specifically, upon historical interactions drawn from the training data $\Set{D}_{\text{train}}$, a CF model is learned to generalize well on future interactions from the test data $\Set{D}_{\text{test}}$, considering underlying popularity distribution shift:
\begin{gather}\label{equ:test-likelihood}
    \theta^{*}=\argmax_{\theta}\log{P_{\text{test}}(Y|X)},
\end{gather}
where $\theta^{*}$ is the oracle parameters of model, which differs from the estimation $\hat{\theta}$ in Equation \eqref{equ:train-likelihood}; $\Set{D}_{\text{test}}$ remains unknown during the training phase.
Worse still, no access is available to quantify the distribution shift \wrt popularity information between $\Set{D}_{\text{train}}$ and $\Set{D}_{\text{test}}$.
Therefore, it is infeasible to solve this generalization problem, without any assumption.

To make reasonable assumptions on popularity generalization, we derive a fine-grained analysis of $P_{\text{test}}(Y|X)$ and reveal how it is distinct from $P_{\text{train}}(Y|X)$ \wrt popularity information.
Specifically, we can decompose $P_{\text{test}}(Y|X)$ into the following terms via the Bayes theorem:
\begin{align}\label{equ:two-terms}
    P_{\text{test}}(Y|X) &= P_{\text{test}}(Y|Z_{\text{pref}}, Z_{\text{pop}})\nonumber\\
    &=\frac{P_{\text{test}}(Z_{\text{pref}}, Z_{\text{pop}}|Y)\cdot P_{\text{test}}(Y)}{P_{\text{test}}(Z_{\text{pref}}, Z_{\text{pop}})}\nonumber\\
    &\propto\underbrace{P_{\text{test}}(Z_{\text{pop}}|Y)}_{\text{Bias term}}\cdot \underbrace{P_{\text{test}}(Z_{\text{pref}}|Y, Z_{\text{pop}})}_{\text{Entanglement term}}.
\end{align}
The bias term shows the direct effect of popularity information $Z_{\text{pop}}$ on $P_{\text{test}}(Y|X)$, whose comparison with $P_{\text{train}}(Z_{\text{pop}}|Y)$ of $P_{\text{train}}(Y|X)$ frames the certain popularity bias.
Meanwhile, the entanglement term depicts that $Z_{\text{pre}}$ are entangled with $Z_{\text{pop}}$, making the distillation of preference information dependent inherently on the popularity information.
Hence, the key to popularity generalization lies in mitigating the influence of popularity on these terms.

\vspace{5pt}
\noindent\textbf{Limitations of Debiasing Approaches.}
Before introducing our assumptions, we first exhibit two deficiencies of current debiasing approaches as follows:
\begin{itemize}[leftmargin=*]
    \item For the bias term in Equation \eqref{equ:two-terms}, most debiasing methods make an implicit but unrealistic assumption that the popularity information in test is available. For example, the reweighting methods \cite{ips,IPS-C,IPS-CN,Propensity_SVM-Rank,RecSys18_IPS,UBPR} (also known as IPS families) actually use the uniform distribution of popularity as the unbiased test data. Recent efforts require the test knowledge in advance to help the model training, such as leveraging the validation set conforming to the test data to guide the hyperparameter adjustment \cite{DICE,MACR,BRD}, or a small unbiased set to boost the unbiased learning \cite{KDCRec,CausE,AutoDebias}. Although these methods could achieve better performance on certain test distribution, they still suffer from other popularity distribution shifts and get degenerated performance.

    \item For the entanglement term in Equation \eqref{equ:two-terms}, prior studies \cite{PDA,MACR,DecRS} simply hypothesize $P_{\text{test}}(Z_{\text{pref}}|Y, Z_{\text{pop}})=P_{\text{train}}(Z_{\text{pref}}|Y, Z_{\text{pop}})$. That is, the popularity information $Z_{\text{pop}}$ has a stable influence on the preference information $Z_{\text{pref}}$ across different distributions.
    However, the \jn{correlation} between $Z_{\text{pop}}$ and $Z_{\text{pref}}$ is naturally shifting.
    Hence, it is crucial to disentangle the preference information from the popularity information.
\end{itemize}
\section{Methodology}

\begin{figure*}[t]
    \centering
    \includegraphics[width=0.9\linewidth]{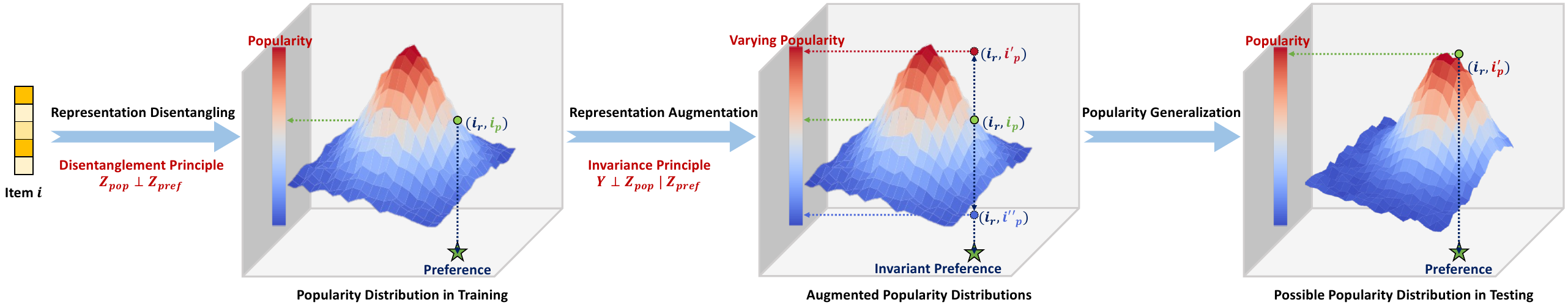}
    \vspace{-10pt}
    \caption{Illustration of item representation derived from InvCF and its popularity generalization.}
    \label{fig:framework}
    \vspace{-10pt}
\end{figure*}
To achieve the popularity generalization, we propose a new learning paradigm, \underline{Inv}ariant \underline{C}ollaborative \underline{F}iltering (InvCF).
Specifically, we begin by introducing two principles:
(1) invariance, which encourages the interaction prediction to be determined by the preference information solely, while invariant to the popularity change;
(2) disentanglement, which enforces the preference and popularity information decomposable and independent from each other.
We then describe our implementation of these principles.


\subsection{Invariant Collaborative Filtering}

Upon the inspection of the bias and entanglement terms in Equation \eqref{equ:two-terms}, we draw inspiration from invariant learning \cite{IRM,REX} and disentangled representation learning \cite{disentangle} to propose two principles.
Next, we elaborate these principles towards popularity generation.

\vspace{5pt}
\noindent\textbf{Invariance Principle.}
By ``invariance'', we conjecture that an ideal CF model should refine the invariant preference information (\eg user preference on item properties) that causally determines the interaction behaviors, regardless of changes in popularity information (\eg user conformity, item popularity).
More formally, this principle is: 
\vspace{-5pt}
\begin{gather}\label{equ:invariance-principle}
   Y \perp Z_{\text{pop}}~|~Z_{\text{pref}},
\end{gather}
where $\perp$ denotes probabilistic independence. It delineates that $Z_{\text{pref}}$ shields $Y$ from the influence of $Z_{\text{pop}}$, making the prediction-making process from $Z_{\text{pref}}$ to $Y$ stable across different $Z_{\text{pop}}$.
Taking movie recommendation as an example, the invariance principle is a lever for looking at users' stable incentives to watch a movie (\eg interest in the director and star aspects), rather than the spurious correlations caused by popularity factors (\eg box office).

\vspace{5pt}
\noindent\textbf{Disentanglement Principle.}
By ``disentanglement'', we mean that the preference and popularity information should be decomposable and independent \cite{disentangle} from each other , so that changing the popularity does not affect the user preference on item properties.
This principle can be formulated as:
\begin{gather}\label{equ:disentanglement-principle}
    Z_{\text{pop}} \perp Z_{\text{pref}}.
\end{gather}
It stipulates that the learning of $Z_{\text{pref}}$ is not susceptible to $Z_{\text{pop}}$.
Considering movie recommendation again, the disentanglement principle makes the popularity of a movie hardly derail a user's interest in the director and star aspects.

\vspace{5pt}
\noindent\textbf{Integration of Two Principles.}
These two principles collaborate with each other to guide the learning of CF models, so as to endow them with powerful prediction and generalization abilities.

\subsection{Implementations of Two Principles}
Here we present how to parameterize these two principles in InvCF.
As illustrated in Figure \ref{fig:framework}, it comprises two additional modules on top of the CF backbone: representation augmentation and disentanglement.
Specifically, the CF backbone is \jn{used} to encode the preference and popularity information as the corresponding representations.
The augmentation module couple a preference representation with diverse popularity representations to approach an invariant prediction, so as to achieve the invariance principle.
Meanwhile, the disentanglement module pursues the independence of preference and popularity representations.
Next, we will elaborate \jn{on} these modules one by one.

\vspace{5pt}
\noindent\textbf{Representation Learning.}
During the training phase, the ground truth of oracle (ideal) preference and popularity information is out of reach, while only the historical user-item interactions are available.
Such an absence motivates us to estimate them in the CF modeling.
Specifically, upon the historical interactions, we employ a CF backbone model to intensify the preference information $Z_{\text{pref}}$ as the representations:
\begin{gather}
    \Mat{u}_{r}, \Mat{i}_{r} = f_{r}(u,i),
\end{gather}
where $f_{r}$ is the CF backbone (\eg MF \cite{BPR}, LightGCN \cite{LightGCN}), termed preference encoder. It takes the ID of user $u$ and item $i$ as the input, and then yields the $d$-dimensional preference representations $\Mat{u}_{r}\in\Space{R}^{d}$ and $\Mat{i}_{r}\in\Space{R}^{d}$, respectively.

Besides the preference encoder $f_{r}$, we hire another popularity encoder $f_{p}$, which has the same architecture to $f_{r}$ but aims to embrace the popularity information $Z_{\text{pop}}$ as the representations:
\begin{gather}
    \Mat{u}_{p}, \Mat{i}_{p} = f_{p}(u,i),
\end{gather}
where $f_{p}$ takes the statistical metrics of popularity as the input (\ie the numbers of interactions that user $u$ and item $i$ are involved in historically) and outputs the $d$-dimensional preference representations $\Mat{u}_{p}\in\Space{R}^{d}$ and $\Mat{i}_{p}\in\Space{R}^{d}$.
It is worth noting that these popularity statistics are treated as categorical features like ID.

As a result, the preference and popularity representations are to estimate $Z_{\text{pref}}$ and $Z_{\text{pop}}$, respectively.
Their combination can parameterize user $u$ and item $i$ as:
\begin{gather}
    \Mat{u} = \Mat{u}_{r}|| \Mat{u}_{p},\quad  \Mat{i} = \Mat{i}_{r}|| \Mat{i}_{p},
\end{gather}
where $||$ denotes the concatenation operation.
To optimize these two encoders, we can adopt the prevalent learning strategy of empirical risk minimization (ERM).
Specifically, the risk function measures the quality of interaction predictions, which can be formulated as BPR loss \cite{BPR} and Softmax loss \cite{ImplicitFeedback,SampledSoftmaxLoss}.
Here we apply Softmax loss $l$ on an observed interaction between user $u$ and item $i$ as: 
\begin{gather}\label{equ:softmax-loss}
    l(\Mat{u}_{r},\Mat{i}_{r})=-\sum_{(u,i)\in\Set{D}_{\text{train}}}\log{\frac{\exp{(s(\Mat{u}_{r},\Mat{i}_{r})/\tau)}}{\sum_{i'\in\Set{N}_{u}\cup\{i\}}\exp{(s(\Mat{u}_{r},\Mat{i}'_{r})/\tau)}}},
\end{gather}
where $s$ is the cosine similarity function;
$\Set{N}_{u}=\{i'|(u,i')\notin\Set{D}_{\text{train}}\}$ is the set of sampled items that $u$ did adopt before, in which $\Mat{i}'_{r}$ is the preference representation of $i'$; $\tau$ is the temperature hyperparameter in softmax \cite{SampledSoftmaxLoss}.
Analogously, we can employ Softmax loss on the popularity representations $\Mat{u}_{p}$ and $\Mat{i}_{p}$.
In a nutshell, the ERM framework minimizes these expected risks:
\begin{gather}
    \Lapl_{\text{rep}}=l(\Mat{u}_{r},\Mat{i}_{r}) + \alpha \cdot l(\Mat{u}_{p},\Mat{i}_{p}),
\end{gather}
which essentially encourages these two representation groups to fit the training interaction and collect the signals relevant to it. 
Here, $\alpha$ is the hyperparameter to control the trade-off between preference and popularity representation learning.
However, solely minimizing the risks over the empirical training distribution suffers from popularity distribution shift \cite{S-DRO}.

\vspace{5pt}
\noindent\textbf{Representation Augmentation.}
To bring forth better generalization \wrt popularity distribution shift, we advocate for the invariance principle in Equation \eqref{equ:invariance-principle}.
Wherein, the relationship between the preference information and the interaction remains stable, regardless of changes in popularity information.
To parameterize this principle, we first devise an augmentation operator on user and item representations, which preserves the estimated preference information, but intervenes the estimated popularity information.
Formally, the operator first collects all popularity representations into two memory banks $\Set{U}_{p}=\{\Mat{u}'_{p}|\forall u'\}$ and $\Set{I}_{p}=\{\Mat{i}'_{p}|\forall i'\}$.
Then it samples a memory $\Mat{u}'_{p}\in\Set{U}_{p}$ to replace $\Mat{u}_{p}$ of $u$, which is combined with $\Mat{u}_{r}$ to create the augmented user $\Mat{u}^{*}$.
Similarly, the augmentation can be performed on $i$ in parallel to construct the augmented item $\Mat{i}^{*}$.
More formally, the augmentation process is:
\begin{gather}
    \Mat{u}^{*}=\Mat{u}_{r}||\Mat{u}'_{p},\quad\Mat{i}^{*}=\Mat{i}_{r}||\Mat{i}'_{p}.
\end{gather}

Having established the augmented representations, we enforce all $(u,i)$'s popularity-intervened versions to hold the consistent discriminative signals about interaction:
\begin{gather}
    \Lapl_{\text{aug}}=\Space{E}_{\Mat{u}'_{p}\in\Set{U}_{p}}l(\Mat{u}^{*},\Mat{i})+\Space{E}_{\Mat{i}'_{p}\in\Set{I}_{p}}l(\Mat{u},\Mat{i}^{*}),
\end{gather}
where $l$ is Softmax loss in Equation \eqref{equ:softmax-loss} but with different inputs.
As a result, it learns to rule out the influence of popularity information, so as to make the preference representations more robust against diversified popularity representations.

\vspace{5pt}
\noindent\textbf{Representation Disentanglement.}
Moreover, we parameterize the disentanglement principle in Equation \eqref{equ:disentanglement-principle} to make the preference and popularity representations independent of each other, so as to assist the invariance principle.
It can be achieved by minimizing a disentanglement regularizer, such as distance correlation \cite{disentangle, DGCF, DICE}, Pearson correlation coefficient \cite{CD2AN}, and Maximum Mean Discrepancy (MMD) \cite{D2Rec} (See Section \ref{sec:related_work}).
Here we minimize the distance correlation $dCor$ between two representation groups:
\begin{gather}
    \Lapl_{\text{dis}}=dCor(\Mat{u}_{r},\Mat{u}_{p}) + dCor(\Mat{i}_{r},\Mat{i}_{p}),
\end{gather}
which takes these representations apart in the latent space.
In conjunction with the invariance principles, it deprives the preference information of mixing the popularity clues.

\vspace{5pt}
\noindent\textbf{Joint Training.}
Overall, we can aggregate all foregoing risks and attain the final objective of InvCF:
\begin{gather}
    \Lapl=\Space{E}_{(x=(u,i),y)\in\Set{D}_{\text{train}}}(\Lapl_{\text{rep}}+\lambda_{1}\cdot\Lapl_{\text{aug}}+\lambda_{2}\cdot\Lapl_{\text{dis}}),
\end{gather}
where $\lambda_{1}$ and $\lambda_{2}$ are the hyperparameters to control the strengths of invariance and disentanglement principles.
Jointly optimizing these risks with these two principles allows the CF backbone (\ie the preference encoder) to focus on the critical cues about users' stable interest in items, regardless of popularity changes.
It endows the CF backbone with better popularity generalization.
In the inference phase, we use the preference representations to make predictions, shielding them from the influence of popularity distribution shifts.

\section{Experiments}
We aim to answer the following research questions:
\begin{itemize}[leftmargin=*]
   \item \textbf{RQ1:} How does InvCF perform compared with other debiasing strategies and popularity generalization baselines?
   \item \textbf{RQ2:} Does InvCF successfully learn popularity-disentangled representations?
   \item \textbf{RQ3:} What are the impacts of the components (\eg disentangling module, augmentation module) on InvCF?
\end{itemize}

\begin{figure}[t]
	\centering
	\vspace{-15pt}
	\subcaptionbox{Training Set\label{fig:yahoo_train}}{
	    \vspace{-6pt}
		\includegraphics[width=0.35\linewidth]{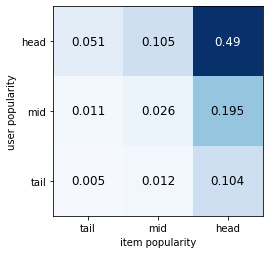}}
	\subcaptionbox{Unbiased Test Set\label{fig:yahoo_test}}{
	    \vspace{-6pt}
		\includegraphics[width=0.465\linewidth]{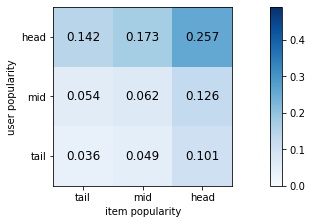}}
	\vspace{-10pt}	
	\caption{An illustration of popularity distribution shifts among different subgroups on Yahoo!R3. 
    Compared to the typical long-tail distribution in the training set, the popularity distribution in the unbiased test set is more evenly distributed.
    See more examples in Appendix \ref{sec:app_datasets}.}
	\label{fig:yahoo}
	\vspace{-15pt}
\end{figure}

\noindent\textbf{Datasets.} 
We conduct extensive experiments on five real-world benchmark datasets (\ie Yahoo!R3 \cite{Yahoo}, Coat \cite{ips}, Douban Movie \cite{Douban}, Meituan \cite{COR}, and Yelp2018 \cite{LightGCN}) and one synthetic dataset (\ie Tencent \cite{Tencent}).
Table \ref{tab:dataset-statistics} provides an overview of the statistics for all datasets, which differ in size, sparsity, domain, and the degree of popularity distribution variations.
Specifically, the popularity shift degree is calculated by KL-divergence between the popularity distribution in the training set and test set, \ie $D_{KL}(P_{train} || P_{test})$, where a higher KL-divergence value indicates a larger shift.
Moreover, to visually demonstrate the distribution varies, we partition the training and test sets into nine disjoint subgroups based on the number of interactions of each item/user: head (the top third), mid (the middle), and tail (the bottom third).
Then we summarize the frequency of interactions across all subgroups.
Figure \ref{fig:yahoo} and Figures \ref{fig:coat}-\ref{fig:meituan} clearly indicate that all datasets exhibit significant popularity distribution shifts between the training and test data.
 
\begin{table}[t]
    \centering
    \vspace{-5pt}
    \caption{Dataset statistics.}
    \label{tab:dataset-statistics}
    \vspace{-10pt}
    \resizebox{1\linewidth}{!}{
    \begin{tabular}{lrrrrrrr}
    \toprule
     & \textbf{Yahoo!R3} & \textbf{Coat} & \textbf{Douban Movie} & \textbf{Meituan}  & \textbf{Tencent} & \textbf{Yelp2018} \\ \midrule
    \#Users & 14,382 & 290 & 36,644 & 67,529  & 95,709  & 4886 \\
    \#Items  & 1,000 & 295 & 22,226  & 29,785 & 41,602  & 4804 \\
    \#Interactions  & 129,748 & 2,776 & 5,397,926 & 2,190,658  &  2,937,228 & 134, 031\\
    Density & 0.0090 &  0.0324 & 0.0066 & 0.0011  & 0.0007 & 0.0057\\ 
    $D_{KL}(P_{train} || P_{test})$  & 0.3561 & 0.1745   & 0.1601& 0.0110  & 0.3639/0.7370/1.1282 & -  \\
   \bottomrule
   \vspace{-5pt}
   \end{tabular}}
\end{table}

\begin{table*}[t]
    \centering
    \caption{The performance comparison on Tencent dataset. The improvement achieved by InvCF is significant ($p$-value $<<$ 0.05).}
    \label{tab:synthetic}
    \vspace{-10pt}
    \resizebox{0.85\linewidth}{!}{
    \begin{tabular}{l|ccc|ccc|ccc|ccc}
    \toprule
     & \multicolumn{3}{c|}{$\gamma$ = 200} & \multicolumn{3}{c|}{$\gamma$ = 10}  & \multicolumn{3}{c|}{$\gamma$ = 2} & \multicolumn{3}{c}{Validation}\\
    \multicolumn{1}{c|}{} & HR & Recall & NDCG & HR  & Recall & NDCG  & HR  & Recall & NDCG & HR  & Recall & NDCG\\\midrule
  MF      & 0.0803 & 0.0292 & 0.0167 & 0.0504 & 0.0188 & 0.0106 & 0.0347 & 0.0132 & 0.0081
          & 0.2537 & 0.0919 & 0.0542\\
+ sam+reg & 0.0811 & 0.0295 & 0.0168 & 0.0525 & 0.0192 & 0.0110 & 0.0352 & 0.0133 
          & 0.0082 
          & 0.2539 & 0.0917 & 0.0543\\
+ IPS-CN  & 0.1299 & 0.0468 & 0.0273 & 0.0894 & 0.0328 & 0.0189 & 0.0656 & \underline{0.0248} 
          & \underline{0.0147} & 0.1702 & 0.0613 & 0.0329\\
+ CausE   & 0.0936 & 0.0340 & 0.0192 & 0.0591 & 0.0212 & 0.0120 & 0.0407 & 0.0153 & 0.0091
          & 0.2461 & 0.0878 & 0.0528\\

+ MACR    & 0.0846 & 0.0301 & 0.0173 & 0.0541 & 0.0203 & 0.0114 & 0.0386 & 0.0149 
          & 0.0089 & 0.2409 & 0.0862 & 0.0488\\
+ sDRO    & \underline{0.1468} & \underline{0.0533} & \underline{0.0311} 
          & \underline{0.0941} & \underline{0.0336}
          & \underline{0.0196} & \underline{0.0664} & 0.0242 & 0.0144 
          & 0.3386 & 0.1318 & 0.0810\\

+ CD$^2$AN    & 0.1409 & 0.0494 & 0.0286 & 0.0852 & 0.0300 & 0.0170 & 0.0569 & 0.0208 & 0.0119 
          & 0.2965 & 0.1108 & 0.0645\\
+ InvCF & \textbf{0.1580}* & \textbf{0.0575}* & \textbf{0.0342}* & \textbf{0.1031}*
          & \textbf{0.0374}* & \textbf{0.0221}* & \textbf{0.0734}* & \textbf{0.0272}* & \textbf{0.0165}* & 0.3230 & 0.1246 & 0.0748 \\\midrule
Imp.\%    & 7.63\% & 7.88\% & 9.97\% & 9.56\% & 11.31\% & 12.76\% & 10.54\% 
          & 9.68\% & 12.24\% & $-$ & $-$ & $-$ \\\midrule\midrule
LightGCN  & 0.1167 & 0.0426 & 0.0240 & 0.0738 & 0.0272 & 0.0151 & 0.0514 & 0.0192 & 0.0114 
          & 0.3018 & 0.1137 & 0.0684 \\
+ sam+reg & 0.1522 & 0.0542 & \underline{0.0307} & \underline{0.1008} 
          & \underline{0.0356} & \underline{0.0198} & \underline{0.0707} & \underline{0.0255} & 0.0141 & 0.3014 & 0.1130 
          & 0.0682 \\
+ IPS-CN  & 0.1316 & 0.0472 & 0.0280 & 0.0874 & 0.0313 & 0.0182 & 0.0644 & 0.0242 
          & \underline{0.0145} & 0.2496 & 0.0920 & 0.0545 \\
+ CausE   & 0.1284 & 0.0469 & 0.0259 & 0.0795 & 0.0289 & 0.0157 & 0.0558 & 0.0210 & 0.0116 
          & 0.2870 & 0.1065 & 0.0659 \\

+ MACR    & 0.1068 & 0.0387 & 0.0208 & 0.0663 & 0.0244 & 0.0131 & 0.0473 & 0.0182 & 0.0101
          & 0.2969 & 0.1122 & 0.0666 \\
+ sDRO    & 0.1455 & 0.0516 & 0.0286 & 0.0857 & 0.0304 & 0.0166 & 0.0552 & 0.0205 & 0.0110  
          & 0.3485 & 0.1374 & 0.0872 \\
+ CD$^2$AN    & \underline{0.1540} & \underline{0.0559} & 0.0305 & 0.0960 & 0.0347 & 0.0185 & 0.0658 & 0.0247 & 0.0134 
          & 0.3594 & 0.1427 & 0.0897 \\
+ InvCF & \textbf{0.1651}* & \textbf{0.0605}* & \textbf{0.0331* }& \textbf{0.1061}* 
          & \textbf{0.0386}* & \textbf{0.0204}* & \textbf{0.0722}* & \textbf{0.0272}* & \textbf{0.0149}* & 0.3611 & 0.1443 & 0.0912 \\\midrule
Imp.\%    & 7.21\% & 8.23\% & 7.82\% & 5.26\% & 8.43\% & 3.03\% & 2.85\% & 7.94\% & 2.76 \% 
          & $-$ & $-$ & $-$ \\\bottomrule
    \end{tabular}}
     \vspace{-10pt}
  \end{table*} 

\noindent\textbf{Test Evaluations.} 
For comprehensive comparisons, three standard test evaluations - unbiased test set \cite{ips,Yahoo}, temporal split test set \cite{PDA,DecRS,sam-reg}, and out-of-distribution test set \cite{COR,InvPref}) - as well as one synthetic evaluation - various popularity distributions as test sets are covered in the experiments. 

\noindent\textbf{Baselines.} 
Two high-performing Collaborative Filtering (CF) models - ID-based (MF \cite{MF}) and graph-based (LightGCN \cite{LightGCN}), are selected as the backbone models being optimized. 
We thoroughly compare InvCF with two backbones and two categories of baselines:
\begin{itemize}[leftmargin=*]
    \item \textbf{Popularity debiasing baselines:} regularization-based \jn{frameworks} (sam+reg \cite{sam-reg}), reweighting-based methods (IPS-CN \cite{IPS-CN}), and causal embedding methods (CausE \cite{CausE}, MACR \cite{MACR}).
    \item \textbf{Popularity domain generalization baselines:} the latest competitive methods CD$^2$AN \cite{CD2AN} which co-trains biased and unbiased models, and sDRO \cite{S-DRO} which improves worst-case performance under distributionally robust optimization framework.
\end{itemize}
See detailed introductions of datasets and baselines in Appendix \ref{sec:app_set}.

\noindent\textbf{Evaluation Metrics.}
We adopt the all-ranking strategy \cite{KricheneR20}, where all items — aside from the positive ones in the training set — are ranked by the CF model for each user. 
Three commonly used metrics—Hit Ratio (HR@$K$), Recall@$K$, and Normalized Discounted Cumulative Gain (NDCG@$K$) — are used to assess the quality of the recommendations, with $K$ being set by default at 20.

\vspace{-10pt}
\textbf{\subsection{Performance Comparison (RQ1)}}

\subsubsection{\textbf{Evaluations on Various Popularity Distributions}}\,\,\,\,

\noindent\textbf{Motivation.}
Many prevalent popularity debiasing techniques concentrate on mitigating bias for single target test distribution \cite{AutoDebias,MACR,ESAM}.
However, the test popularity distributions in real-world recommendation scenarios may be diverse, unpredictable, and unknown. 
We argue that a good CF model is crucial to consistently \jn{yield} a satisfactory performance when dealing with the unidentified popularity distribution shift.
In our settings, the models are identical per method across multiple test sets, and no prior information about the test distribution is provided in advance.

\noindent\textbf{Settings.}
To evaluate the robustness of InvCF and baselines over various popularity distributions, we randomly select three long-tailed subsets of interactions as the test sets to mimic the popularity distribution shift, with each subset containing 10\% interactions.
Concretely, we first rank items in descending order and divide them into 50 groups according to their popularity.
Then, for the $i$-th group, $N_i = N_0 \cdot \gamma^{-\frac{i-1}{49}}$ interactions are sampled out to generate the test set. 
Here, $N_0$ is the maximum number of interactions among all the groups in \jn{the} test, and $\gamma$ describes the long-tail degree.
A smaller $\gamma$ indicates a stronger distribution shift and a more uniform popularity distribution compared to the training data.
Besides, the test splits for the validation set are equally long-tailed as the train set, \ie randomly split the remaining interactions into training, and validation sets (60\% : 10\%).
Figure \ref{fig:tencent} shows the popularity distributions of the training and three test sets on Tencent.

\noindent\textbf{Results.}
Table \ref{tab:synthetic} reports the comparison of performance on all the baselines with different levels of long-tail degree.  
The best\jn{\-}performing methods per test are bold and starred, while the strongest baselines are underlined; Imp.\% measures the relative improvements of InvCF over the strongest baselines. 
We observe that:
\begin{itemize}[leftmargin=*]
    \item \textbf{InvCF consistently and significantly outperforms the state-of-the-art baselines in terms of all metrics across all popularity distributions.}
    Specifically, InvCF achieves remarkable improvements compared to the best baselines by 10.2\% and 5.9\% \wrt NDCG@20 on average over the MF and LightGCN backbones, respectively.
    We attribute the robustness of InvCF to \jn{distilling} the invariant preference information, regardless of changes in popularity distributions.
    \item \textbf{Entangled models behave in an unstable manner when the popularity distribution shifts. }
    Compared to CD$^2$AN and sDRO, popularity debiasing methods show a limited enhancement of recommendation quality over backbone models, reflecting a lack of generalization capability when the target prior is unknown.
    When $\gamma =2$, as expected, IPS-CN performs the second best as its unbiased test defaults to a uniform distribution.
    With a closer look at performance on \jn{the} validation set, debiasing models tend to gain improvements in \jn{tests} by sacrificing the fitness on training data. 
    In contrast, benefiting from no implicit assumption on the test data, CD$^2$AN, sDRO and our InvCF can substantially boost the performance over backbone models by a large margin.
    These observations serve to corroborate our study in Section \ref{sec:prelimimary} that entangled representations and strict \jn{requirements} of prior information are two deficiencies of current debiasing approaches.  
\end{itemize}

\begin{figure*}[ht]
	\centering
	\subcaptionbox{BPR loss\label{fig:head_bpr}}{
	    \vspace{-6pt}
		\includegraphics[width=0.175\linewidth]{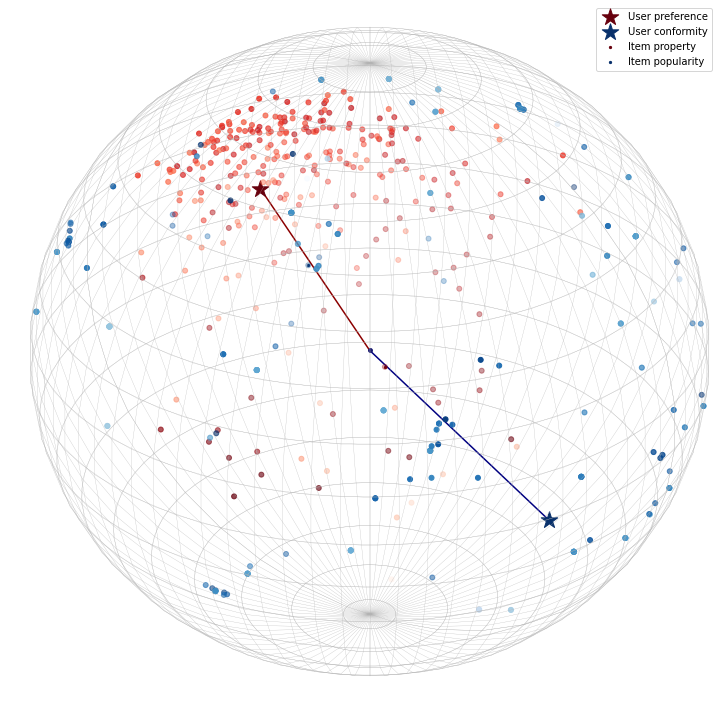}}
	\subcaptionbox{Softmax loss \label{fig:head_softmax}}{
	    \vspace{-6pt}
		\includegraphics[width=0.175\linewidth]{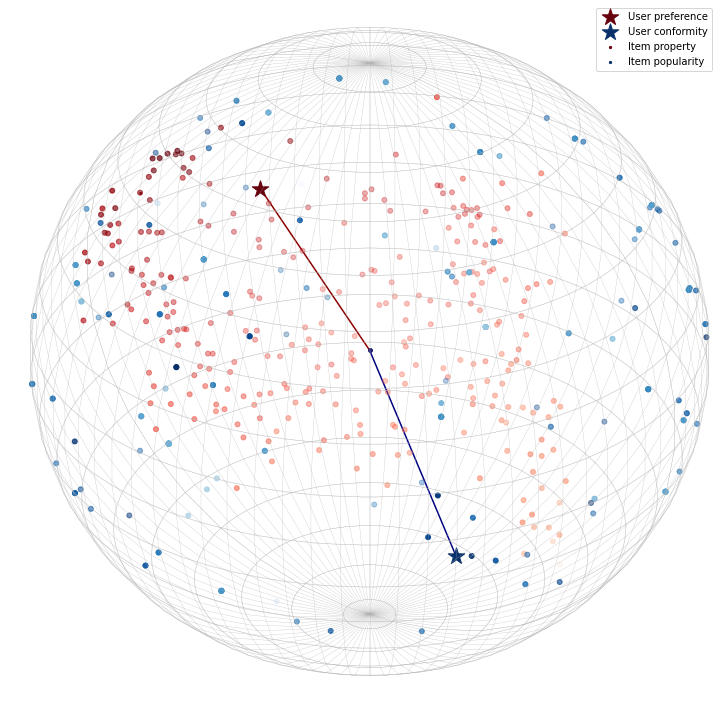}}
    \subcaptionbox{InvCF-i\label{fig:head_noaug}}{
	    \vspace{-6pt}
		\includegraphics[width=0.175\linewidth]{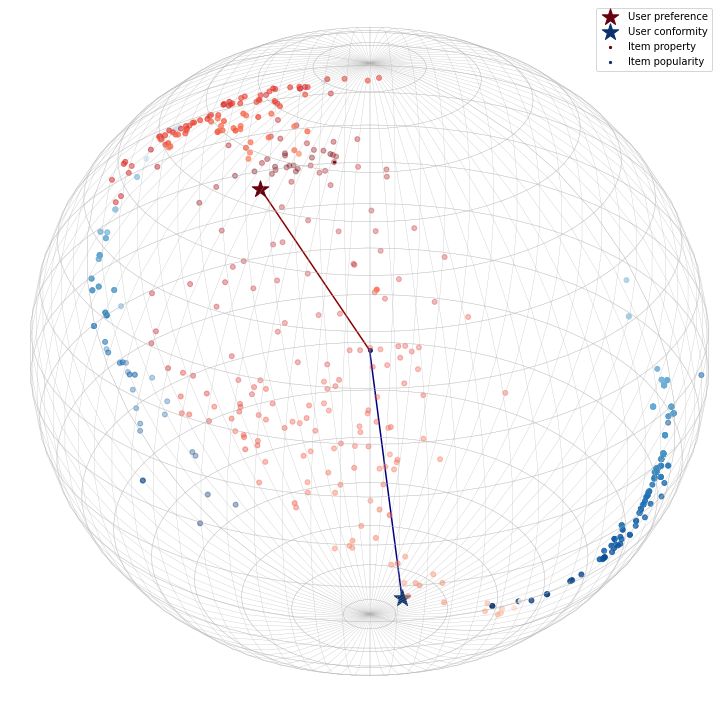}}
	\subcaptionbox{InvCF\label{fig:head_InvCF}}{
	    \vspace{-6pt}
		\includegraphics[width=0.243\linewidth]{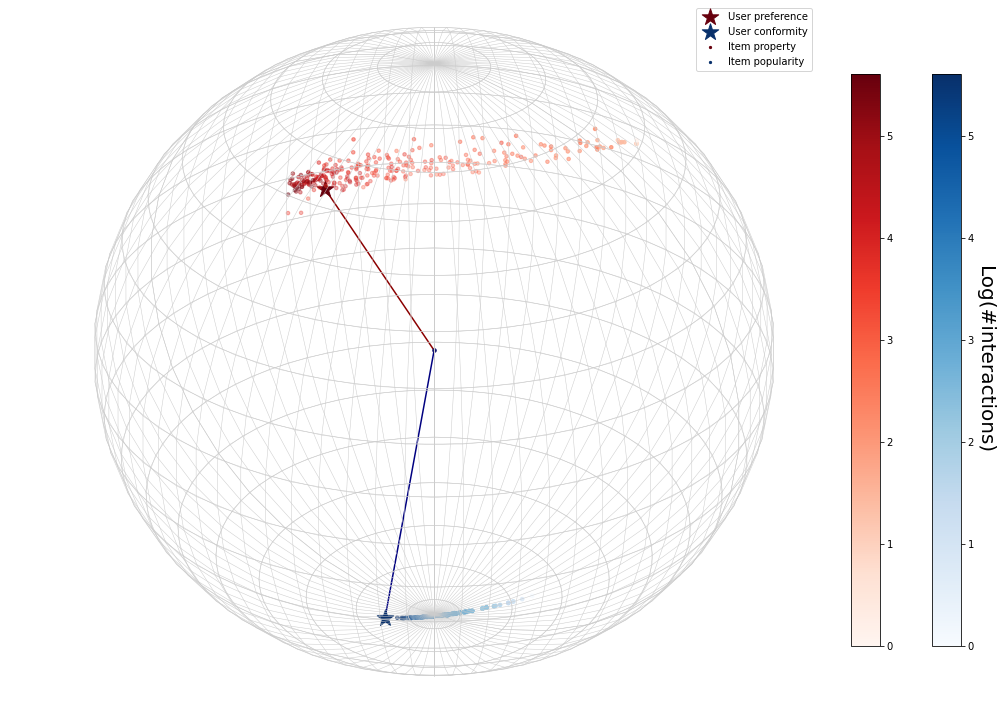}}
	\vspace{-10pt}
	\caption{3D Visualizations of item representations learned by MF backbone model on Yelp2018.
	Subfigures (a-d) showcase the preference and property representations of the identical head user as red and blue stars, respectively.
	In each subfigure, representations of the head user's all historical items are projected on the unit sphere.
	The brightness of the color indicates the popularity degree, while red and blue dots denote preference and popularity representations of items, respectively.
	More visualization results can be found in Appendix \ref{sec:app_3d}.}
	\label{fig:3d_viz}
	\vspace{-5pt}
\end{figure*}

\begin{table}[t]
    \centering
    \caption{The performance comparison on Yahoo!R3 and Coat. }
    \label{tab:yahoo_coat}
    \vspace{-10pt}
    \resizebox{\linewidth}{!}{
    \begin{tabular}{l|cc|cc|cc|cc}
    \toprule
     & \multicolumn{4}{c|}{Yahoo!R3} & \multicolumn{4}{c}{Coat}  \\ 
     \multicolumn{1}{c|}{} & \multicolumn{2}{c|}{MF} & \multicolumn{2}{c|}{LightGCN} & \multicolumn{2}{c|}{MF} & \multicolumn{2}{c}{LightGCN} \\ 
    \multicolumn{1}{c|}{} & Recall & NDCG  & Recall & NDCG & Recall & NDCG  & Recall & NDCG \\\midrule
     Backbone & 0.1063 & 0.0476  & 0.1478 & 0.0686   & 0.0741 & 0.0361 
     & 0.2658 & 0.1574\\
    + sam+reg & 0.1198 & 0.0548  & 0.1498 & \underline{0.0693} & 0.2303 & 0.1869 & \underline{0.2659} & 0.3569\\
    + IPS-CN  & 0.1081 & 0.0487  & 0.1331 & 0.0612  & 0.1700 & 0.1377 & 0.2474 & 0.1771\\
    + CausE   & 0.1252 & 0.0573 & 0.1490 & 0.0693  & 0.2004 & 0.1713 & 0.2479 & 0.1689 \\
    + MACR    & 0.1243 & 0.0539 & \underline{0.1499} & 0.0691  & 0.0798 & 0.0358 & 0.0939 & 0.0584 \\
    + sDRO    & 0.1390 & 0.0661 & 0.1426 & 0.0660 & 0.2012 & 0.1767 & 0.2415 & \underline{0.1790}  \\
   + CD$^2$AN & \underline{0.1451} & \underline{0.0690} & 0.1397 & 0.0638  & \underline{0.2325} & \underline{0.1885} & 0.2245 & 0.1708 \\
    + InvCF   & \textbf{0.1566}* & \textbf{0.0732}*  & \textbf{0.1515}* & \textbf{0.0718}* & \textbf{0.2672}* & \textbf{0.2059}* & \textbf{0.2686}* & \textbf{0.1819}*  \\\midrule
    Imp.\%    & 7.93\% & 6.09\% &  1.07\% & 3.61\% & 14.43\% & 9.23\%  & 1.01\% & 1.62\% \\\bottomrule
    \end{tabular}}
    \vspace{-12pt}
  \end{table}

\subsubsection{\textbf{Evaluations on Unbiased Test Sets}}\qquad\qquad\qquad\qquad\qquad\,\,
\noindent\textbf{Motivation.}
Offline evaluation on collaborative filtering is challenging because of the missing-not-at-random condition in real-world recommender systems. 
Unbiased evaluation, where its test set is composed of items selected at random rather than by users, is considered an ideal offline test for eliminating the impact of CF models. 
Here we conduct experiments on widely used missing-complete-at-random datasets: Yahoo!R3 and Coat.

\noindent\textbf{Results.}
As Table \ref{tab:yahoo_coat} depicts that InvCF steadily superior over all baselines \wrt all metrics on Yahoo!R3 and Coat.
Specifically, compared to the state-of-the-art baseline, it achieves substantial gains on the MF backbones over Yahoo!R3 and Coat in terms of Recall@20 by 7.93\% and 14.43\%, respectively.   
In contrast, baselines perform unstably across datasets.
Consistent with our study, this validates that InvCF successfully parameterizes the invariance and disentanglement principle, resulting a popularity-disentangled representations.

\subsubsection{\textbf{Evaluations on Temporal split and Out-of-distribution Test Sets}}\qquad\qquad\qquad\qquad\qquad\qquad\qquad\qquad\qquad\qquad\qquad\qquad\,\,

\begin{table}[t]
    \centering
    \caption{The performance comparison on Douban Movie. }
    \label{tab:douban}
    \vspace{-10pt}
    \resizebox{0.83\linewidth}{!}{
    \begin{tabular}{l|ccc|ccc}
    \toprule
      \multicolumn{1}{c|}{} & \multicolumn{3}{c|}{MF} & \multicolumn{3}{c}{LightGCN}  \\
    \multicolumn{1}{c|}{} & HR & Recall & NDCG &  HR & Recall & NDCG 
    \\\midrule
     Backbone     & 0.3509 & 0.0289 & 0.0552 & 0.3569 & 0.0294 & 0.0558\\
     + sam+reg & 0.3584 & 0.0303 & 0.0577 & 0.3569 & 0.0307 & 0.0575 \\
    + IPS-CN  & 0.2844 & 0.0216 & 0.0403 & 0.3213 & 0.0268 & 0.0522 \\
    + CausE  & 0.3587 & 0.0300 & 0.0579 & 0.3640 & 0.0310 & 0.0589 \\
    + MACR    & 0.3559 & \underline{0.0303} & \underline{0.0584} & 0.3620 & 0.0310 & 0.0595\\
    + sDRO   & \underline{0.3670} & 0.0303  & 0.0553 & 0.3707 & \underline{0.0324} & \underline{0.0618} \\
    + CD$^2$AN    & 0.3602 & 0.0296	& 0.0553 & \underline{0.3770}	&0.0327 &	0.0604 \\
    + InvCF & \textbf{0.3757}* & \textbf{0.0321}* & \textbf{0.0604}* & \textbf{0.3897}* & \textbf{0.0343}* & \textbf{0.0635}*\\\midrule
    Imp.\%    & 2.37\% & 5.94\% & 3.42\% & 3.37\% & 5.86\% & 2.75\%\\\bottomrule
    \end{tabular}}
    \vspace{-12pt}
  \end{table}

\noindent\textbf{Motivation.}
In real-world applications, popularity distribution dynamically changes over time.
For a comprehensive comparison, we take two time-related evaluations into consideration.
On Douban Movie, we divide the historical interactions into the training, validation, and test sets according to the timestamps (7:1:2).
On Meituan, following the settings in \cite{COR}, the user interactions during weekdays are regarded as the training (60\%) and validation (10\%) sets, while user purchases during the weekend are used as the test set (30\%).
See popularity distributions of each sets in Appendix \ref{sec:app_datasets}.

\begin{table}[t]
    \centering
    \caption{The performance comparison on Meituan. }
    \label{tab:meituan}
    \vspace{-10pt}
    \resizebox{0.9\linewidth}{!}{
    \begin{tabular}{l|ccc|ccc}
    \toprule
      \multicolumn{1}{c|}{} & \multicolumn{3}{c|}{MF} & \multicolumn{3}{c}{LightGCN}  \\
    \multicolumn{1}{c|}{} & HR & Recall & NDCG &  HR & Recall & NDCG 
    \\\midrule
     Backbone     & 0.5490 & 0.2343 & 0.2250  & 0.5760 & 0.2545 & 0.2574 \\
     + sam+reg & 0.5518 & 0.2358 & 0.2272 & 0.5797 & 0.2590 & \underline{0.2664}\\
    + IPS-CN  & 0.5311 & 0.2207 & 0.2046 & 0.5592 & 0.2448 & 0.2449\\
    + CausE  & 0.5665 & 0.2456 & 0.2409 & 0.5849 & \underline{0.2615} & 0.2619 \\
    + MACR    & 0.5583 & 0.2368 & 0.2129  & 0.5779 & 0.2587 & 0.2522\\
    + sDRO   & \underline{0.5922} & \underline{0.2648} & \underline{0.3002}  & \underline{0.5929} & 0.2279 & 0.2289\\
    + CD$^2$AN    & 0.5914 & 0.2466  & 0.2664 & 0.5751	& 0.2509 & 	0.2634\\
    + InvCF & \textbf{0.5954}* & \textbf{0.2780}* & \textbf{0.3073}* & \textbf{0.6132}* & \textbf{0.2642}* & \textbf{0.2778}*\\\midrule
    Imp.\%    & 0.54\% &  4.98\% & 2.37\%  & 3.42\% & 1.03 \% & 4.28\% \\\bottomrule
    \end{tabular}}
    \vspace{-12pt}
  \end{table}

\noindent\textbf{Results.}
Tables \ref{tab:douban} and \ref{tab:meituan} clearly show, InvCF yields a consistent boost compared to the SOTA baselines. 
This indicates that InvCF endows the backbone models with better robustness and generalization ability against changes in the popularity distribution that are caused by time.
Considering \jn{the} empirical success of InvCF on test evaluations, we believe that InvCF provides a promising research direction to cope with popularity distribution shifts.

\vspace{-10pt}
\textbf{\subsection{Visualizations of Representations (RQ2)}}

\begin{figure}[t]
    \centering
    \vspace{-10pt}
    \includegraphics[width=0.55\linewidth]{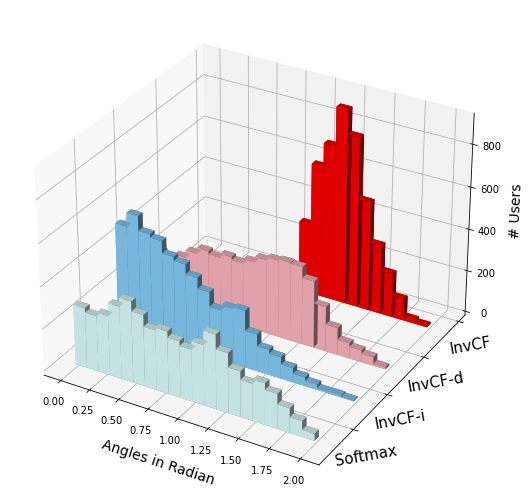}
    \vspace{-10pt}
    \caption{Distribution of angles between preference and popularity representations.}
    \label{fig:angle}
    \vspace{-10pt}
\end{figure}

To visualize the latent representation space and evaluate the effectiveness of two principles in InvCF, we train toy recommenders on Yelp2018 using the MF backbone whose embedding size is three.
We compare InvCF with two variants: InvCF-i, which disables the augmentation module, and InvCF-d, which deactivates the disentangling module.
In Figure \ref{fig:3d_viz}, 3-dimensional normalized item preference and popularity representations under different CF models (\ie MF+BPR, MF+Softmax \cite{SSM}, MF+InvCF-i, MF+InvCF) are illustrated on a 3D unit sphere.
In Figure \ref{fig:angle}, we further summarize the distributions of angles between the preference and popularity representations learned by different CF models. 
We observe that:
\begin{itemize}[leftmargin = *]
    \item \textbf{Item preference and popularity representations learned by BPR and SSM are chaotically distributed and difficult to distinguish}, as shown by Figures \ref{fig:3d_viz} and \ref{fig:3d_viz_appendix}.
    This validates that BPR and SSM extract entangled representations, further resulting in learning a suboptimal latent space.
    \item \textbf{Item preference and popularity representations learned by InvCF are disentangled and discriminative.}
    Figure \ref{fig:angle} reveals that angles learned by InvCF between preference and popularity representations tend to be more concentrated, leading to stable latent spaces.
    This clearly shows that InvCF not only effectively learns popularity-disentangled representations but also successfully distills the invariant features that causally determine the interactions.
    We attribute this breakthrough to  representation-level augmentation and disentangling modules.
\end{itemize}

\vspace{-13pt}
\textbf{\subsection{Study on InvCF (RQ3)}}

\noindent\textbf{Ablation Study.}
Jointly analyzing the results in Figures \ref{fig:3d_viz}, \ref{fig:angle}, \ref{fig:3d_viz_appendix}, and Table \ref{tab:ablation}, we observe that:
\begin{itemize}[leftmargin=*]
    \item \textbf{From a performance perspective, invariance and disentanglement principles are critical for InvCF and indispensable for one another.}
    In particular, InvCF-i and InvCF-d are susceptible to degrade the recommendation quality as shown in Table \ref{tab:ablation}.
    Furthermore, with additional regularization terms, InvCF surprisingly converges on fewer training epochs, highlighting the fact that lacking any principle may undermine the model learning.
    \item \textbf{In terms of representations, InvCF incorporates disentangling and augmentation modules to decouple the invariant features and spurious popularity, further improving generalization ability.}
    Precisely, from Softmax to InvCF-i in Figure \ref{fig:3d_viz}, the preference and popularity representations present a clear boundary, demonstrating the efficiency of the disentangling module.
    However, as Figure \ref{fig:angle} shows, the overall angle distributions of InvCF-i, InvCF-d, and Softmax barely differ from one another.
    We ascribe the limited difference to \jn{employing} one module alone, rather than two modules in cooperation, thus failing to acquire stable and high-quality feature spaces.  
    Compared to InvCF-i and InvCF-d, InvCF drives \jn{an} impressive breakthrough in representation learning, implying the cooperation of disentangling and augmentation modules is integral.
\end{itemize}

\begin{table}[t]
    \centering
    \caption{Ablation Study on Yahoo!R3 dataset.}
    \label{tab:ablation}
    \vspace{-10pt}
    \resizebox{0.8\linewidth}{!}{
    \begin{tabular}{l|cccc}
    \toprule
     & HR & Recall & NDCG & $\#$epoch \\ \midrule
    Softmax & \multicolumn{1}{l}{0.2224} & \multicolumn{1}{l}{0.1470} & \multicolumn{1}{l}{0.0688} & 264.5 \\
    InvCF-i & $0.2195 ^{\color{blue}-1.30 \%}$ & $0.1457 ^{\color{blue}-0.88 \%}$ & $0.0701 ^{\color{red}+1.89 \%}$ & 304.8 \\
    InvCF-d & $0.2241 ^{\color{red}+0.76 \%}$ & $0.1464 ^{\color{blue}-0.41 \%}$  & $0.0696 ^{\color{red}+1.16 \%}$ & 419.1\\ 
    InvCF  & $\textbf{0.2333}^{\color{red}+4.90 \%}$ & $\textbf{0.1566}^{\color{red}+6.53 \%}$ & 
    $\textbf{0.0732}^{\color{red}+6.40 \%}$ & 164.2\\
   \bottomrule
    \end{tabular}}
    \vspace{-13pt}
\end{table}

\noindent\textbf{Effect of Disentangling/Augmentation Module.} Assembling different discrepancy regularizers in the disentangling module will lead to performance fluctuations (See Table \ref{tab:discrepancy}), while InvCF might be insensitive to various augmentation strategies (See Table \ref{tab:aug}).
\section{Related Work} \label{sec:related_work}
\textbf{Popularity Debiasing in recommender.} Leading popularity debiasing approaches can roughly fall into three research lines:
\begin{itemize} [leftmargin=*]
    \item \textbf{Regularization-based frameworks} ~\cite{sam-reg, ESAM, ALS+Reg, PC_Reg} regulate the trade-off between accuracy and coverage with additional penalty terms.
    ESAM ~\cite{ESAM} introduces 
    self-training regularizers to handle the missing of target labels. 
    ALS+Reg ~\cite{ALS+Reg} leverages intra-list diversity (ILD) as the regularization. 
    \item \textbf{Re-weighting methods} ~\cite{ips, IPS-CN, BRD, DR, AutoDebias} re-weight the loss of training instances by inverse propensity scores. 
    Recent studies explore learning unbiased propensity estimators (AutoDebias \cite{AutoDebias}, BRD ~\cite{BRD}) instead of directly adopting the observed frequencies (IPS-CN \cite{IPS-CN}) to lower the variance of propensities.
    
    \item \textbf{Causal embedding learning} ~\cite{MACR, CausE, DICE, PDA, KDCRec, DecRS, CauSeR} utilizes counterfactual inference to mitigate the effect of item popularity.
    DecRS ~\cite{DecRS} uses backdoor adjustment to eliminate the effect of imbalanced item group distribution. 
    PDA ~\cite{PDA} adopts do-calculus ~\cite{Causality} to remove the confounding popularity bias.
\end{itemize}
\textbf{Out-of-distribution (OOD) Generalization.} 
To devise stable models and address the problem of distribution shifts, 
s-DRO ~\cite{S-DRO} adopts Distributionally Robust Optimization (DRO) framework, CD$^{2}$AN ~\cite{CD2AN} disentangles item property representations from popularity under co-training networks, 
while BC Loss \cite{bc_loss} incorporates bias-aware margins to achieve better generalization ability.
Another line of research incorporates causal inference and discovery into OOD generalization (CausPref \cite{CausPref}, COR \cite{COR}). \\
\noindent \textbf{Disentangled Representation Learning in CF.} 
To disentangle user or item intents into a finer granularity, current CF models assemble different discrepancy metrics including distance correlation (DGCF \cite{DGCF}, DICE \cite{DICE}), Pearson Correlation Coefficient (CD$^2$AN \cite{CD2AN}), Maximum Mean Discrepancy (D2Rec \cite{D2Rec}) and $L_1$, $L_2$ normalization (DICE \cite{DICE}). 
Another line of research utilizes Variational Auto Encoders (VAE) to factorize latent features \cite{beta-VAE, MacridVAE, DisenHAN}.

\noindent \textbf{Data Augmentation in CF.} 
Data augmentation is a popular strategy in recommender systems to deal with cold-start ~\cite{AR-CF} or data sparsity problems~\cite{AugCF}. 
However, semantic augmentation in CF remains stagnant.
To the best of our knowledge, we are among the first to undertake representation level data augmentation to enhance the popularity generalization ability.
We believe that this work provides a promising research line to deal with the OOD problem and will shed light on future work.
\section{Conclusion}
Leading popularity debiasing methods in collaborative filtering are still far from resolving the recommender's vulnerability to popularity distribution shift.
In this work, we proposed a novel learning strategy, InvCF, that extracts invariant and popularity-disentangled features to enhance popularity generalization ability.
Grounded by extensive test evaluations and real-world visualization studies, InvCF steadily outperforms the SOTA baselines by learning disentangled representation spaces. 
A worthwhile direction for future work is to extend InvCF to handle generic distribution shifts in recommender systems. 
We believe that InvCF will inspire research to incorporate latent space disentanglement and augmentation.


\begin{acks}
This research is supported by the Sea-NExT Joint Lab, the National Natural Science Foundation of China (9227010114), the University Synergy Innovation Program of Anhui Province (GXXT-2022-040), and CCCD Key Lab of the Ministry of Culture and Tourism.
\end{acks}

\bibliographystyle{ACM-Reference-Format}
\bibliography{2023_www}

\appendix
\clearpage
\section{Experimental Settings}  \label{sec:app_set}
\subsection{Datasets.} \label{sec:app_datasets}
\begin{figure}[t]
	\centering
	\subcaptionbox{Training Set\label{fig:tencent_train}}{
	    \vspace{-6pt}
		\includegraphics[width=0.36\linewidth]{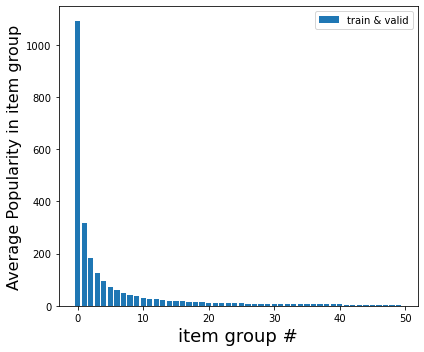}}
	\subcaptionbox{Synthetic Test Set\label{fig:tencent_test}}{
	    \vspace{-6pt}
		\includegraphics[width=0.36\linewidth]{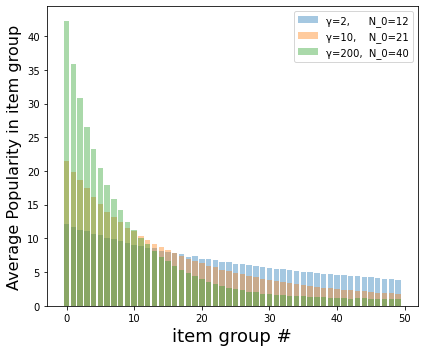}}
	\vspace{-10pt}	
	\caption{Popularity distributions on Tencent.}
	\label{fig:tencent}
	\vspace{-10pt}
\end{figure}

We conduct experiments on both real-world benchmark datasets and one synthetic dataset. 
\begin{enumerate}
     \item \textbf{Yahoo!R3 \cite{Yahoo}} \& \textbf{Coat \cite{ips}}: These two datasets are obtained from the music and coat recommendation services. Both Yahoo!R3 and Coat are specially designed to evaluate on unbiased settings. The training data, considered as a normal biased dataset, contains ratings for items selected by users. The testing data is collected from an online survey, where each user has to rate on randomly selected items. 
    \item \textbf{Douban Movie \cite{Douban}}:  This dataset is collected  from a popular movie review website Douban in China. We split it based on the temporal splitting strategy \cite{MengMMO20}. 
    \item \textbf{Meituan \cite{COR}}: Meituan is a public food recommendation dataset. The shifts of consumption levels from weekdays to weekends causes popularity drift in items.
    \item \textbf{Yelp2018 \cite{LightGCN}}: This dataset is a subset of Yelp's businesses, reviews, and user data. We randomly sample 10\% of interactions from the original dataset to do the 3D visualization.
    \item \textbf{Tencent \cite{Tencent}}: The original dataset is collected from Tencent short-video platform. We generate three testing datasets to explore model performance on dataset with different distribution shifts. The popularity distributions are shown in Figure \ref{fig:tencent}. 
\end{enumerate}

\begin{figure}[t]
	\centering
	\subcaptionbox{Training Set\label{fig:coat_train}}{
	    \vspace{-6pt}
		\includegraphics[width=0.336\linewidth]{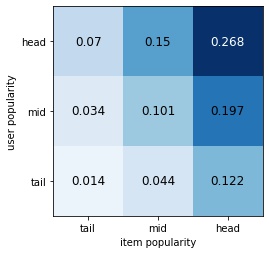}}
	\subcaptionbox{Unbiased Test Set\label{fig:coat_test}}{
	    \vspace{-6pt}
		\includegraphics[width=0.447\linewidth]{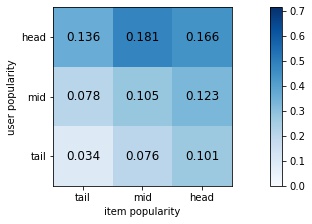}}
	\vspace{-10pt}	
	\caption{Popularity distribution on Coat.}
	\label{fig:coat}
	\vspace{-10pt}
\end{figure}

\begin{figure}[t]
	\centering
	\subcaptionbox{Training Set\label{fig:douban_train}}{
	    \vspace{-6pt}
		\includegraphics[width=0.336\linewidth]{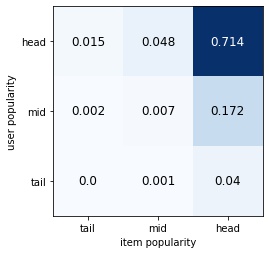}}
	\subcaptionbox{Temporal Split Test Set\label{fig:douban_test}}{
	    \vspace{-6pt}
		\includegraphics[width=0.447\linewidth]{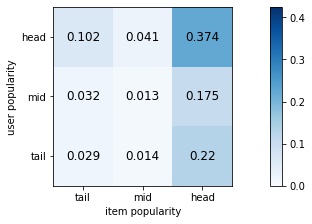}}
	\vspace{-10pt}	
	\caption{Popularity distributions on Douban Movie.}
	\label{fig:douban}
	\vspace{-10pt}
\end{figure}

\begin{figure}[t]
	\centering
	\subcaptionbox{Training Set (Weekday)\label{fig:meituan_train}}{
	    \vspace{-6pt}
		\includegraphics[width=0.336\linewidth]{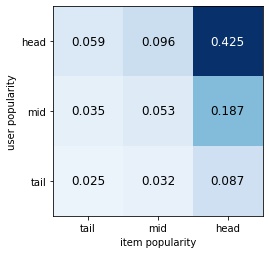}}
	\subcaptionbox{Test Set (Weekend)\label{fig:meituan_test}}{
	    \vspace{-6pt}
		\includegraphics[width=0.447\linewidth]{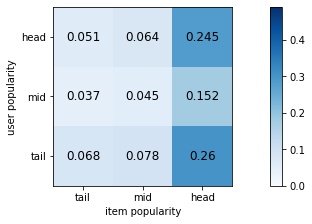}}
	\vspace{-10pt}	
	\caption{Popularity distributions on Meituan.}
	\label{fig:meituan}
	\vspace{-10pt}
\end{figure}

Figures \ref{fig:yahoo}, \ref{fig:coat}-\ref{fig:tencent} show the popularity distribution shift from training to testing.
In the stage of pre-processing data, explicit feedback (Yahoo!R3, Coat) are converted into implicit feedback. We treat the items rated with four or higher scores (five in total) as positive feedback and the remaining as negative feedback. Following the standard 10-core setting \cite{PDA,KGAT}, we filter out items and users with less than ten interactions for all five datasets.

\begin{figure*}[t]
	\centering
	\subcaptionbox{BPR loss (head)\label{fig:head_bpr_1}}{
	    \vspace{-6pt}
		\includegraphics[width=0.21\linewidth]{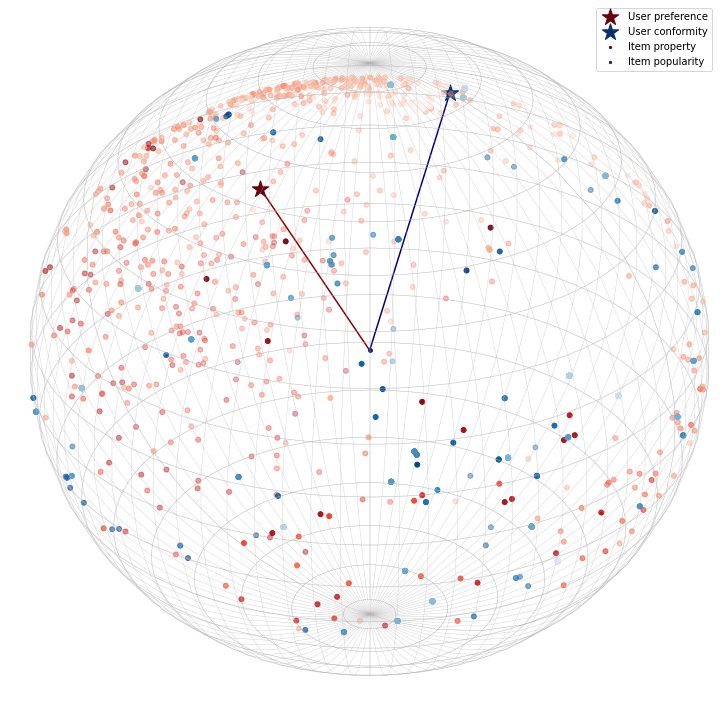}}
	\subcaptionbox{Softmax loss (head)\label{fig:head_softmax}}{
	    \vspace{-6pt}
		\includegraphics[width=0.21\linewidth]{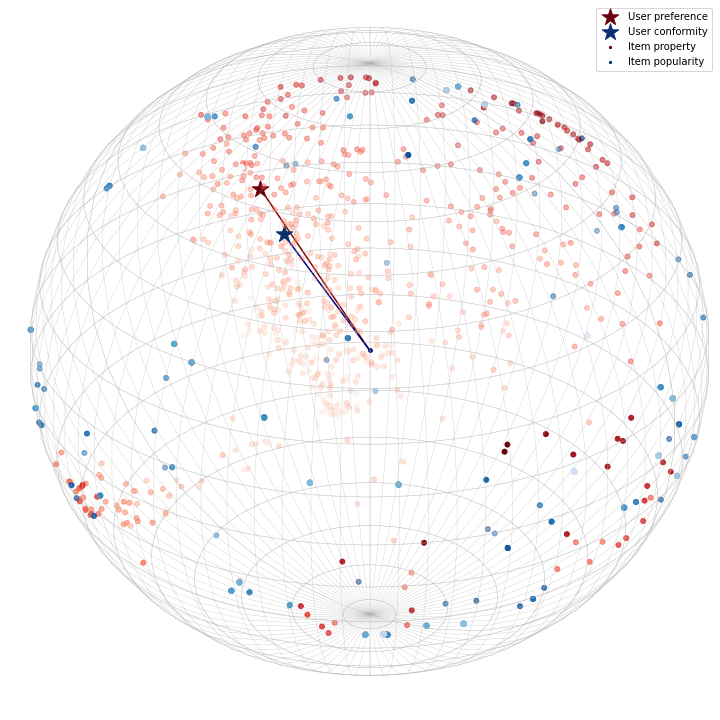}}
    \subcaptionbox{InvCF-i (head) \label{fig:head_InvCF-i_1}}{
	    \vspace{-6pt}
		\includegraphics[width=0.21\linewidth]{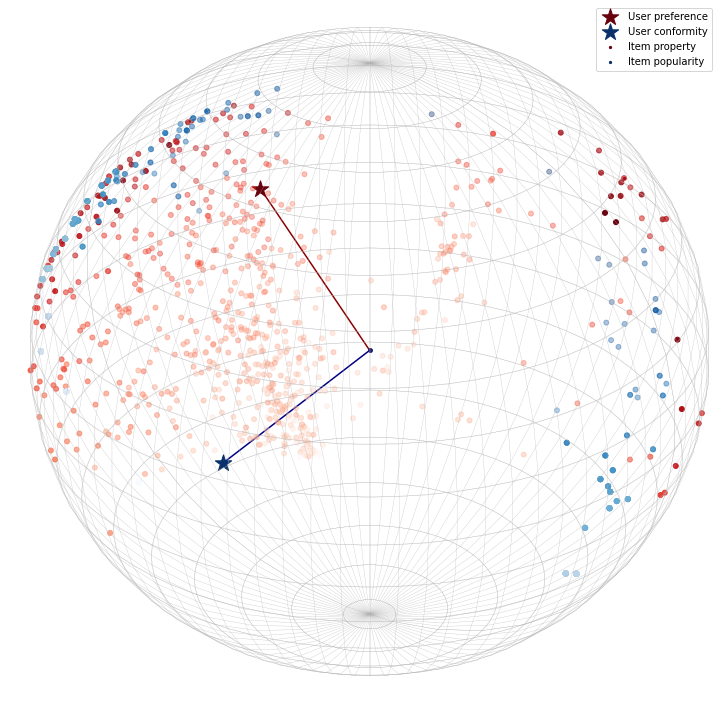}}
	\subcaptionbox{InvCF (head)\label{fig:head_InvCF_1}}{
	    \vspace{-6pt}
		\includegraphics[width=0.28\linewidth]{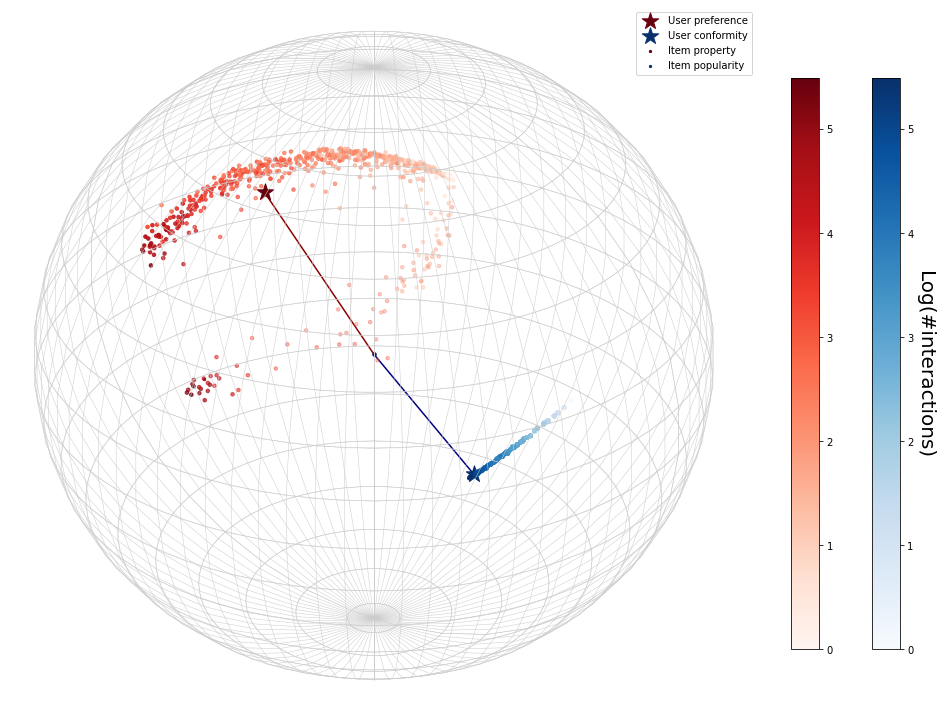}}

	\subcaptionbox{BPR loss (tail)\label{fig:tail_bpr}}{
	    \vspace{-6pt}
		\includegraphics[width=0.21\linewidth]{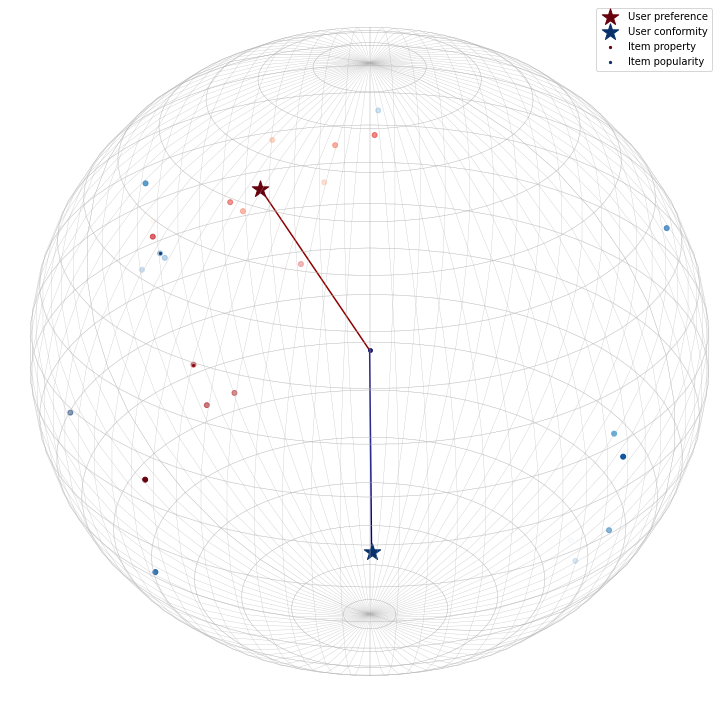}}
	\subcaptionbox{Softmax loss (tail)\label{fig:tail_softmax}}{
	    \vspace{-6pt}
		\includegraphics[width=0.21\linewidth]{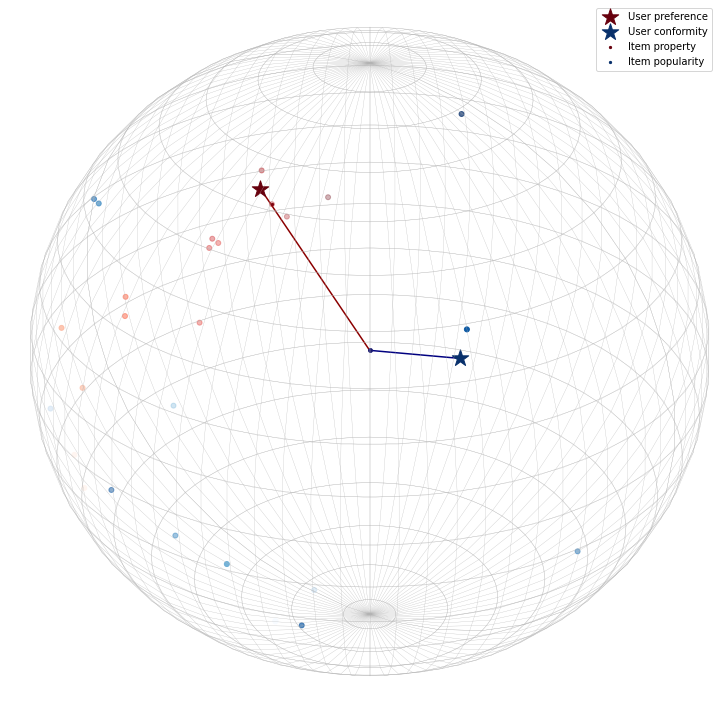}}
    \subcaptionbox{InvCF-i (tail)\label{fig:tail_noaug}}{
	    \vspace{-6pt}
		\includegraphics[width=0.21\linewidth]{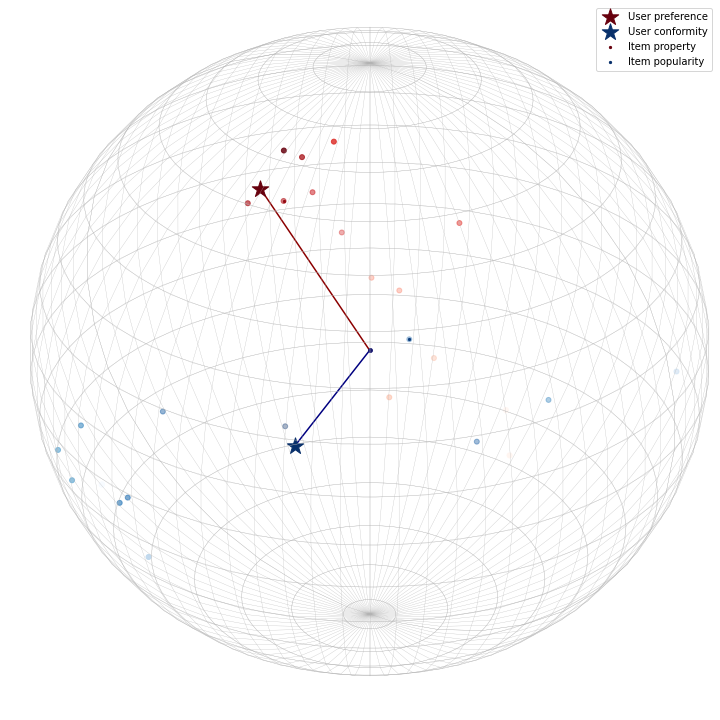}}
	\subcaptionbox{InvCF (tail)\label{fig:tail_InvCF}}{
	    \vspace{-6pt}
		\includegraphics[width=0.29\linewidth]{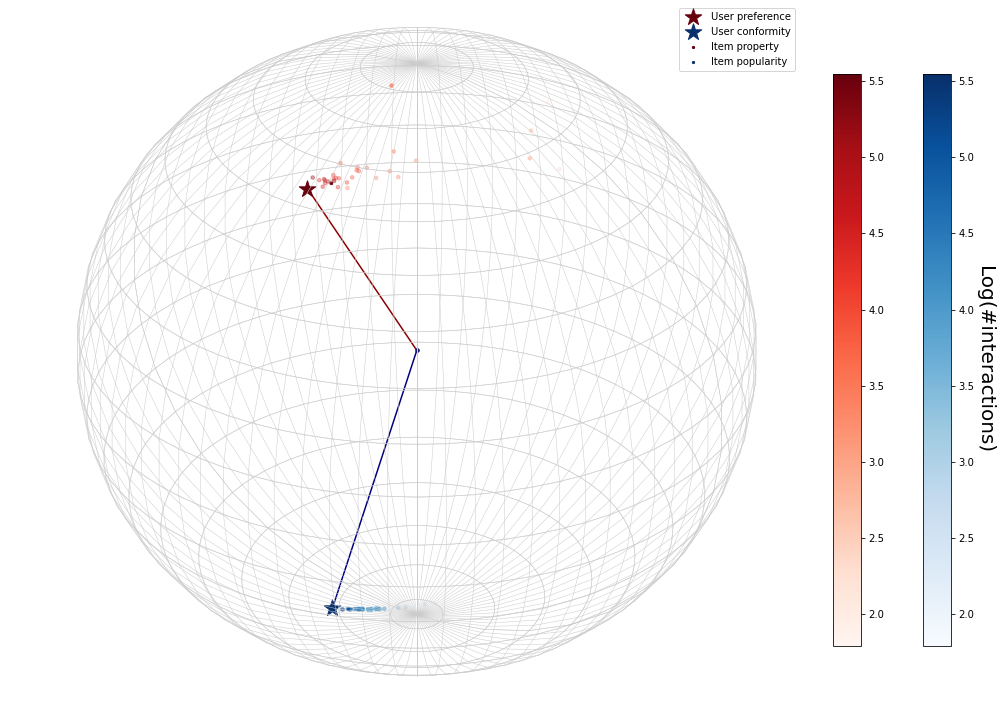}}
    \vspace{-10pt}
	\caption{Additional 3D Visualizations of item representations learned by MF backbone model on Yelp2018.}
	\label{fig:3d_viz_appendix}
\end{figure*}

\subsection{Baselines} \label{sec:app_baselines}
We compare with popular debiasing strategies in various research lines: sam+reg \cite{sam-reg}, IPS-CN \cite{IPS-CN}, and CausE \cite{CausE}, MACR \cite{MACR}. We also compare with domain generalization baselines: CD$^2$AN \cite{CD2AN} and s-DRO \cite{S-DRO}.

\begin{itemize}
    \item \textbf{sam+reg} \cite{sam-reg}: This method comprises two parts: training examples mining (sam) to balance the distribution of observed and unobserved items, and regularized optimization (reg) to minimize biased correlations between predicted user-item relevance and item popularity. 
    \item \textbf{IPS-CN} \cite{IPS-CN}: IPS \cite{ips} re-weights each training instance with item popularity to eliminate popularity bias. IPC-CN adds normalization on plain IPS to achieve lower variance.
    \item \textbf{CausE} \cite{CausE}: This method leverages a small unbiased dataset to simulate the training process under a fully random recommendation policy. 
    \item \textbf{MACR} \cite{MACR}: This method 
    assigns popularity bias to the causal effects of item popularity on the prediction scores. To this end, it introduces two additional modules to capture the effects of item popularity and user conformity and injects the results into the final prediction scores.
    \item \textbf{CD$^2$AN} \cite{CD2AN} This model uses Pearson coefficient correlation to disentangle item property representations from item popularity representation and introduces additional unexposed items to align item popularity distributions between hot and long-tail items.
    \item \textbf{s-DRO} \cite{S-DRO}: This model adds streaming optimization improvement to the Distributionally Robust Optimization (DRO) framework to mitigate the amplification of Empirical Risk Minimization (ERM) on popularity bias.
\end{itemize}

\subsection{Parameter Settings and Training Cost}
Table \ref{tab:parameter} and \ref{tab:parameter_baseline} show the parameter settings for InvCF and baselines, respectively. 
We further record the training cost of InvCF compared to selected baselines in Table \ref{tab:elapse_time}.

\begin{table}
    \centering
    \caption{Training cost on Yahoo!R3 (seconds per epoch/in total). }
    \vspace{-10pt}
    \label{tab:elapse_time}
    \resizebox{\linewidth}{!}{
    \begin{tabular}{l|rrrrrrrr}
    \toprule
     & Backbone & +sam+reg & +IPS-CN & +CausE & +MACR & sDRO & CD$^{2}$AN & InvCF\\ \midrule
    MF & 1.5 / 230 & 1.5 / 74 & 1.3 / 538 & 1.6 / 86 & 1.2 / 119 & 2.0 / 918 & 2.5 / 960  & 3.3 / 541 \\
    LightGCN & 1.8 / 232 & 1.8 / 232 & 1.7 / 66 & 2.0 / 328 & 1.9 / 369  & 4.7 /66 & 2.3 / 378 & 4.5 / 738\\ \bottomrule
    \end{tabular}}
\end{table}





\begin{table}
    \centering
    \caption{Model architectures and hyperparameters for InvCF.}
    \label{tab:parameter}
    \vspace{-10pt}
    \resizebox{\columnwidth}{!}{
    \begin{tabular}{l|c|c|c|c|c|c|c}
    \toprule
     \multicolumn{1}{c|}{} & \multicolumn{7}{c}{InvCF hyper-parameters} \\\midrule
     
    \multicolumn{1}{c|}{} & $\alpha$ & $\lambda_1$ & $\lambda_2$ & $\tau$ & lr & batch size & No. negative samples \\\midrule
    \textbf{MF} \\\midrule
    Yahoo!R3 & 1e-4 & 1e-5 & 1e-3  &  0.15 & 5e-4 & 1024 & 128 \\\midrule
    Coat & 1e-3 & 1e-6 & 1e-2  & 0.09 & 5e-4 & 1024 & 64 \\\midrule
    Douban  & 1e-2 & 1e-5 & 1  & 0.13 & 5e-4 & 1024 & 128 \\\midrule
    Meituan & 1 & 1e-4 & 0  & 0.03 & 5e-4 & 1024 & 128 \\\midrule
    Tencent & 1 & 1 & 1e-8  & 0.09 & 5e-4 & 1024 & 128 \\\midrule
   
    \textbf{LightGCN} \\\midrule
    Yahoo!R3 & 1 & 1e-1 & 1e-7  &  0.18 & 5e-4 & 1024 & 64 \\\midrule
    Coat & 1 & 1e-4 & 0  & 0.95 & 5e-4 & 64 & inbatch \\\midrule
    Douban  & 1e-4 & 1e-2 & 1e-2  & 0.13 & 5e-4 & 1024 & 128 \\\midrule
    Meituan & 1e-1 & 1e-6 & 0  & 0.03 & 5e-4 & 1024 & inbatch \\\midrule
    Tencent & 1e-2 & 1e-2 & 1e-6  & 0.17 & 5e-4 & 1024 & inbatch \\\midrule
    \end{tabular}}
    \vspace{-5pt}
\end{table}

\begin{table}
    \centering
    \caption{Hyper-parameters search spaces for baselines.}
    \label{tab:parameter_baseline}
    \vspace{-10pt}
    \resizebox{\columnwidth}{!}{
    \begin{tabular}{l|l}
    \toprule
     \multicolumn{1}{c|}{} & \multicolumn{1}{c}{Hyper-paramete space} \\\midrule
     
    \textbf{MF $\&$ LightGCN} & \makecell{lr = 5e-4, batch size $\sim $ \{64, 128, 256, 512, 1024, 2048\} \\
    No. negative samples $\sim$  \{128, 256, 512, inbatch\} }\\\midrule
    \textbf{sam-reg} & Log(rweight) $\sim $ \{-1,-2,-3,-4,-5,-6, -7, -8\}  \\\midrule
    \textbf{CausE} & Log($cf\_pen$) $\sim $ \{-1,-2,-3,-4,-5,-6, -7, -8\} \\\midrule
    \textbf{MACR} & Log($c$) $\sim $ \{-1,-2,-3,-4,-5,-6, -7, -8\} \\\midrule
    \textbf{CD$^2$AN} & \makecell{$\tau \sim $  \{0.03, 0.05, 0.07, 0.11, 0.13, 0.15, 0.17\}\\
     Log($\lambda_1$) $\sim $ \{-1,-2,-3,-4,-5\},  Log($\lambda_2$) $\sim $ \{-1,-2,-3,-4,-5\}} \\\midrule
    \textbf{sDRO} & \makecell{$\tau \sim $  \{0.03, 0.05, 0.07, 0.11, 0.13, 0.15, 0.17\}\\
    $t_1, t_2 \sim $  \{0.1, 0.2, 0.3, 0.4, 0.5, 0.6, 0.7, 0.8, 0.9\}\\
     Log($dro\_temperature$) $\sim $ \{-1,-2,-3\},  Log($streaming\_lr$) $\sim $ \{0, -1,-2\}}\\\midrule
    \end{tabular}}
\end{table}

\begin{table}
    \centering
    \caption{Ablation Study on Different Discrepancy Metrics.}
    \label{tab:discrepancy}
    \vspace{-10pt}
    \resizebox{0.8\linewidth}{!}{
    \begin{tabular}{l|cccc}
    \toprule
    & \multicolumn{4}{c}{NDCG@20} \\ 
     & $\gamma = 200$ & $\gamma = 10$ & $\gamma = 2$ & Validation\\ \midrule
    $dCor$ & \multicolumn{1}{l}{0.0342} & \multicolumn{1}{l}{0.0221} & \multicolumn{1}{l}{0.0165} & 0.0748 \\
    MMD & $0.0355 ^{\color{red}+3.80 \%}$ & $0.0241 ^{\color{red}+9.05 \%}$ & $0.0182 ^{\color{red}+10.30 \%}$ & 0.0475 \\
    $L_2$ & $0.0279 ^{\color{blue}+18.42 \%}$ & $0.0178 ^{\color{blue}-24.16 \%}$  & $0.0135 ^{\color{blue}-18.18 \%}$ & 0.0863\\ 
   \bottomrule
    \end{tabular}}
\end{table}

\begin{table}
    \centering
    \caption{Ablation Study on Different Data Augmentation Approaches.}
    \label{tab:aug}
    \vspace{-10pt}
    \resizebox{\linewidth}{!}{
    \begin{tabular}{l|ccc}
    \toprule
    & \multicolumn{3}{c}{NDCG@20} \\ 
     & $\gamma = 200$ & $\gamma = 10$ & $\gamma = 2$ \\ \midrule
     \makecell{\textbf{Random Permutation} : randomly shuffle the orders of \\ popularity embeddings inside the current batch.}
     & 0.0342 & 0.0221 & 0.0165\\\midrule
     \makecell{\textbf{Head Group} : for each user/item, randomly sample a\\ popularity embedding from head group users/items.}
     & 0.0335 & 0.0215 & 0.0163 \\\midrule
     \makecell{\textbf{Tail Group} : for each user/item, randomly sample a\\ popularity embedding from tail group users/items.}
     & 0.0332 & 0.0214 & 0.0161  \\\midrule
    \makecell{\textbf{Different Groups} : $i.e.$ for a user in head group, randomly \\sample a popularity embedding from mid and tail groups.} & 0.0333 & 0.0217 & 0.0162 \\
   \bottomrule
    \end{tabular}}
    \vspace{-10pt}
\end{table}

\subsection{Additional Experiments} \label{sec:app_3d}
 \noindent\textbf{3D Visualizations}
 We equally divide users into three subgroups: head, mid, and tail, according to their interaction frequencies. We then select the most popular one from the head group and another from the tail group to illustrate the feature representations of their interacted items in 3D space in Figure \ref{fig:3d_viz_appendix}.

\noindent\textbf{Effect of Disentangling Module.}
Table \ref{tab:discrepancy} shows the performance comparison among deploying different discrepancy metrics in the disentangling module, including Maximum Mean Discrepancy (MMD), distance correlation and $L_2$ normalization.

\noindent\textbf{Effect of Augmentation Module.}
Table \ref{tab:aug} illustrates the performance comparison among various augmentation strategies.

\end{document}